\documentclass[aps,pra,amsmath,amssymb,twocolumn,superscriptaddress,showpacs]{revtex4-1}
\usepackage{graphicx}
\usepackage{url}

\begin{document}

\title{Phase noise of self-sustained optomechanical oscillators}

\author{King Yan Fong}
\affiliation{Department of Electrical Engineering, Yale University, New Haven, CT 06511, USA}
\author{Menno Poot}
\affiliation{Department of Electrical Engineering, Yale University, New Haven, CT 06511, USA}
\author{Xu Han}
\affiliation{Department of Electrical Engineering, Yale University, New Haven, CT 06511, USA}
\author{Hong X. Tang}
\affiliation{Department of Electrical Engineering, Yale University, New Haven, CT 06511, USA}

\date{\today}

\begin{abstract}
In this paper we present a theory that predicts the phase noise characteristics of self-sustained optomechanical oscillators. By treating the cavity optomechanical system as a feedback loop consisting of an optical cavity and a mechanical resonator, we analytically derive the transfer functions relating the amplitude/phase noise of all the relevant dynamical quantities from the quantum Langevin equations, and obtain a closed-form expressions for the phase noise spectral densities contributed from thermomechanical noise, photon shot noise, and low-frequency technical laser noise. We numerically calculate the phase noise for various situations and perform a sample calculation for an experimentally demonstrated system. We also show that the presented model reduces to the well-known Leeson's phase noise model when the amplitude noise and the amplitude/phase noise inter-transfers are ignored.
\end{abstract}


\maketitle

\section{Introduction}

Self-sustained oscillation (also known as auto-oscillation) is of great interest in both fundamental studies and technological applications \cite{Jenkins_PhysRep_2013}. In fact, operations of today's electronics heavily depend on self-sustained oscillators for time keeping and frequency control purposes. For example, quartz oscillators are routinely used to generate clock signals for digital circuits \cite{Vig_Quartz}, while RF and microwave oscillators are used to provide stable time and frequency references for the operations of wireless communications, radar, and remote sensing systems \cite{Ludwig_Book_2000, Pozar_Book_2012}.

A self-sustained oscillator can be modeled by a simple feedback loop consisting of two loop components: a frequency selective element and an amplifier \cite{Pozar_Book_2012, Rubiola_Book_2010}. When the loop has positive feedback and a loop gain greater than one, the signal is reinforced after each round-trip and as a result the system becomes unstable and self-oscillates. In electronic oscillators, electrical as well as mechanical resonators are commonly used as the frequency selective element and the signal is amplified by electrical amplifiers. In the case of optoelectronic oscillators, the signal can as well be amplified through the transduction process of electro-optical modulation and optical detection \cite{Maleki_JOSAB_1996, Zhou_IEEEIFCS_2007}. Self-oscillation can also be compactly realized in a cavity optomechanical system, where an optical cavity is coupled with a mechanical resonator through the optical force \cite{Vahala_OE_2007,Marquardt_Arxiv_2013}. In such system, the mechanical motion is encoded in the light circulating inside the cavity, which, after a phase delay, exerts an optical force back to the mechanical resonator. When the input laser is blue detuned from the cavity resonance, the optical force provides a positive feedback and amplify the mechanical motion, which can be strong enough to compensate the mechanical damping and drive the resonator into self-sustained oscillation.

Such optomechanical oscillations were first demonstrated by Vahala group using micro-toroid structures \cite{Vahala_OE_2005, Vahala_PRA_2006, Vahala_PRA_2008}, which then stimulated a series of research works realizing optomechanical oscillations using different device designs and materials, such as flexural beams \cite{Kippenberg_NatPhys_2009, Tang_NatNano_2011}, micro-disk resonators \cite{Painter_PRL_2009, Lin_OE_2012}, micro-wheel resonators \cite{Bhave_OE_2011, Bhave_IEEEMEMS_2012, Nguyen_IEEEMEMS_2013, Tang_APL_2012}, micro-spheres \cite{Carmon_PRL_2009}, capillaries \cite{Carmon_NatCommun_2013}, photonic crystals \cite{Painter_Nature_2009}. Technical aspects of these oscillations such as injection-locking \cite{Vahala_APL_2008, Wong_APL_2013}, photonic/RF down-conversion \cite{Vahala_IEEEPTL_2008, Hossein-Zadeh_JLT_2013}, synchronization \cite{Lipson_PRL_2012, Tang_PRL_2013} and mass sensing applications \cite{Hossein-Zadeh_OE_2013} have also been explored. These demonstrations have oscillating frequencies ranging from radio-frequency up to microwave X-band, effective mass from nanogram down to femtogram, and threshold operating power as low as micro-Watt, showcasing the optomechanical oscillator as a promising candidate for scalable, CMOS-compatible, micron-scale, low-threshold power and low phase noise oscillators. While several theoretical studies have been devoted to investigate the multistability \cite{Girvin_PRL_2006}, amplitude noise suppression \cite{Armour_PRL_2010}, limit-cycle behavior \cite{Tang_PRA_2012}, non-classical characteristics \cite{Marquardt_PRL_2012, Hammerer_PRX_2014} of the system and a few phenomenological models have been developed to model the phase noise behavior \cite{Vahala_APL_2006, Vahala_PRA_2008, Bhave_IFCS_2010, Maleki_OE_2012}, however, there is still lack of theoretical understanding which can deterministically predict the phase noise spectra, which is the uttermost important figure of merit for quantifying the performance of an oscillator.

In this paper, we develop a theory the predicts the phase noise of self-sustained optomechanical oscillators. We treat the optomechanical system as a feedback loop \cite{Stamper-Kurn_PRA_2012, vanderZant_PhysRep_2012}, with the optical cavity acts as an amplifier and the mechanical resonator as the frequency selective element. Instead of linearizing the equation of motion around a static solution, we perform the perturbation around a limit-cycle solution and solve for the small-signal transfer functions for each loop component. Using the transfer function approach, we are able to derive analytic expressions for the phase noise spectrum contributed from the thermomechanical noise, photon shot noise and low-frequency technical laser noise. We examine the various factors that influence the phase noise and apply the theory to calculate the phase noise spectrum for the optomechanical oscillator demonstrated in Refs. \cite{Vahala_OE_2005, Vahala_PRA_2006, Vahala_PRA_2008}. We also show that, when the amplitude noise and the amplitude/phase noise inter-transfers are ignored, the present model reduces to the well-known Leeson's model \cite{Leeson_IEEE_1966}, which is widely used for modeling the phase noise of oscillators in the engineering community. This paper does not only addresses the practical questions concerning phase noise performance, such as the fundamental limit of the phase noise that can be attained in optomechanical oscillators, but also provides a theoretical framework for studying cavity optomechanical systems with large oscillation amplitude. 

The outline of the paper is as follow. Section \ref{sec:formalism} presents the theoretical framework of the phase noise analysis for an optomechanical system. First, Sec. \ref{subsec:QLE} examines the coupled quantum Langevin equations that govern the dynamic of the optomechanical system. Sec. \ref{subsec:smallsignal} presents the small-signal analysis and clarify the meaning of phase noise in different frequency scales. After that, in Sec. \ref{subsec:tfunc}, the transfer functions relating the amplitude and phase noise of all the relevant dynamical quantities are derived. The derived expressions are then applied in Sec. \ref{sec:results} to solve for the limit-cycle solution (Sec. \ref{subsec:limit_cycle}) and closed-loop noise response (Sec. \ref{subsec:noise_response}). The phase noise contribution from three sources: thermomechanical noise (Sec. \ref{subsec:phasenoise_thermal}), photon shot noise (Sec. \ref{subsec:phasenoise_shot}) and low-frequency technical laser noise (Sec. \ref{subsec:phasenoise_tech}) are then examined. In Sec. \ref{subsec:sample_calc} the theory is applied to calculate the phase noise spectrum for an experimentally demonstrated system reported in Refs. \cite{Vahala_OE_2005, Vahala_PRA_2006, Vahala_PRA_2008}. Sec. \ref{subsec:leeson_model} compares the present phase noise theory with the well-known Leeson's model \cite{Leeson_IEEE_1966}, and finally Sec. \ref{sec:conclusion} concludes the paper.

\section{Formalism}\label{sec:formalism}

\subsection{Quantum Langevin equations of motion}\label{subsec:QLE}

\begin{figure}[t]
\centering
\includegraphics[width=8cm]{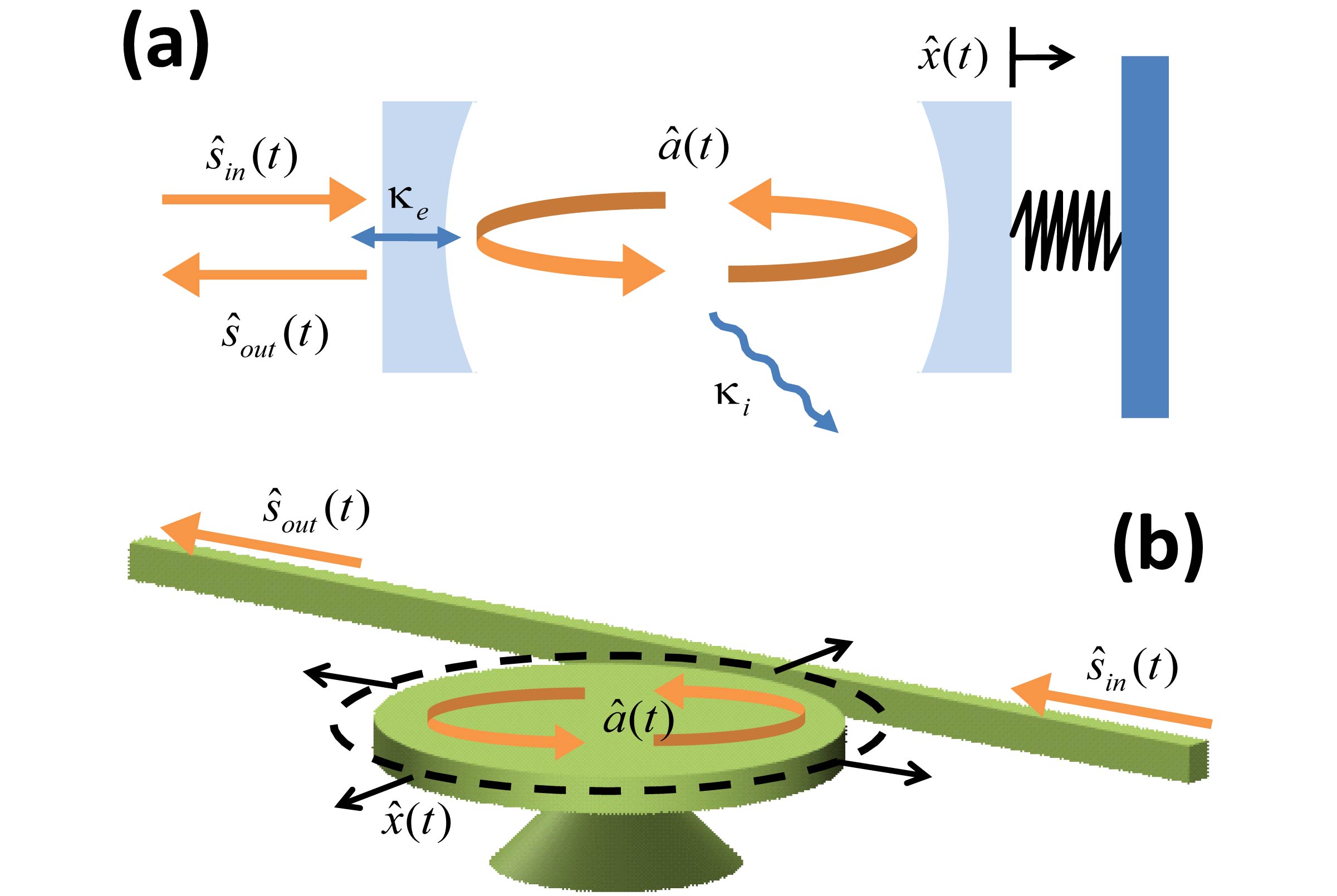}
\caption{ (a) Schematics of a Fabry-P\'{e}rot cavity with a movable mirror. The output field $\hat{s}_{out}(t)$ represents the reflected light from the cavity (b) Schematics of a micro-disk resonator with movable boundary. The output field $\hat{s}_{out}(t)$ represents the transmitted light in the thru-port. }
\label{fig:setup}
\end{figure}

Figure~\ref{fig:setup} shows two typical configurations of optomechanical systems. The first system is in a setting of free-space optics consisting of a Fabry-P\'{e}rot cavity with a movable mirror. An input laser beam is used to excite the cavity. The second system is an integrated photonic resonator with movable boundary. A side-coupling waveguide is used to couple light into the cavity. For the integrated photonic approach, other configurations such as a flexural beam evanescently coupled to a micro-disk \cite{Kippenberg_NatPhys_2009}, or photonic crystal cavity with movable boundaries \cite{Painter_Nature_2009} can also be used. While the optomechanical system can be realized in a wide variety of configurations, its dynamics can be described by the same Hamiltonian \cite{Marquardt_Arxiv_2013}
\begin{equation}\label{eq:Hamiltonian}
\begin{aligned}
\hat{H} =& \hbar\Omega_O\hat{a}^\dagger\hat{a} +\frac{1}{4}\hbar\Omega_M(\hat{x}^2+\hat{p}^2)
-\hbar g\hat{x}\hat{a}^\dagger\hat{a} \\
&+ i\hbar\sqrt{2\kappa_e} (\hat{a}^\dagger\hat{s}_{in}e^{-i\Omega_L t} -\hat{a}\hat{s}_{in}^\dagger e^{i\Omega_L t}) + \hat{H}_{diss} ~.
\end{aligned}
\end{equation}
The first term describes an optical mode at frequency $\Omega_O$ with $\hat{a}$ and $\hat{a}^\dagger$ as the ladder operators. The second terms describes a mechanical mode at frequency $\Omega_M$, and $\hat{x}$ and $\hat{p}$ are respectively the displacement and momentum operators normalized by the zero-point fluctuations $x_{zpf}\equiv\sqrt{\hbar/2m_e\Omega_M}$ and $p_{zpf}\equiv\sqrt{\hbar m_e\Omega_M/2}$, where $m_e$ is the effective mass of the resonator. The third term describes the coupling between the optical and mechanical mode, in which the photon number $\hat{a}^\dagger\hat{a}$ and mechanical displacement $\hat{x}$ are coupled through the coupling rate $g$. $g$ is also called vacuum optomechanical coupling rate since it has a physical meaning of change of optical mode frequency caused by the mechanical zero-point fluctuation. The fourth term represents the optical drive by an input optical mode with field strength of $\hat{s}_{in}$ at frequency $\Omega_L$. This input optical mode can be one of the laser beam modes in free-space optics (Fig. \ref{fig:setup}~(a)), or the waveguide mode in integrated photonics (Fig. \ref{fig:setup}~(b)). $\kappa_e$ represents the coupling rate between this input mode and the cavity mode. The last term accounts for the dissipation due to the link to the external bath.

The quantum Langevin equations describing the dynamics of $\hat{x}$ and $\hat{a}$ (in the rotating frame $e^{-i\Omega_L t}$) can be obtained from the Hamiltonian in Eq. \eqref{eq:Hamiltonian} as \cite{Marquardt_Arxiv_2013}
\begin{eqnarray}
\dot{\hat{a}} = -\kappa\hat{a}+i\Delta\hat{a} +ig\hat{x}\hat{a} +\sqrt{2\kappa_e}\hat{s}_{in} +\sqrt{2\kappa_i}\delta\hat{s}_{vac} \label{eq:QLE_opt} \\
\ddot{\hat{x}} +2\gamma\dot{\hat{x}} +\Omega_M^2\hat{x} =2\Omega_M(g\hat{a}^\dagger\hat{a} +\sqrt{2\gamma}\delta\hat{f}_{th}) ~. \label{eq:QLE_mech}
\end{eqnarray}
Eq. \eqref{eq:QLE_opt} describes the dynamics of the cavity mode. $\Delta$ is the cavity detuning defined as $\Delta=\Omega_L-\Omega_O$. The cavity is coupled to the environment (input mode) by a coupling rate $\kappa_i$ ($\kappa_e$), where the subscript $i$($e$) represents the intrinsic (external) nature of the coupling. Coupling to these two channels gives rise to a total dissipation rate of $\kappa=\kappa_e+\kappa_i$. Note that here $\kappa$ is defined as the half width at half maximum (HWHM) of the cavity response. (Some references define $\kappa$ as the full width at half maximum, see for example Ref. \cite{Marquardt_Arxiv_2013}.) It is related to the cavity quality factor by $Q_O=\Omega_O/2\kappa$. The input field $\hat{s}_{in}$ can be split into two parts $\hat{s}_{in}=s_{in}+\delta\hat{s}_{in}$, with the scalar part $s_{in}$ represents the laser field strength and the operator part $\delta\hat{s}_{in}$ represents the input laser noise. $\delta\hat{s}_{vac}$ represents the quantum vacuum fluctuation due to the coupling to the environment and has noise correlators given by
\begin{equation}\label{eq:noise_corr_vac}
\begin{aligned}
\langle\delta\hat{s}_{vac}(t)\delta\hat{s}_{vac}^\dagger(t^\prime)\rangle &= \delta(t-t^\prime) \\
\langle\delta\hat{s}_{vac}^\dagger(t)\delta\hat{s}_{vac}(t^\prime)\rangle &= 0 \\
\langle\delta\hat{s}_{vac}(t)\delta\hat{s}_{vac}(t^\prime)\rangle &= 0 \\
\langle\delta\hat{s}_{vac}^\dagger(t)\delta\hat{s}_{vac}^\dagger(t^\prime)\rangle &= 0 ~.
\end{aligned}
\end{equation}
When the laser noise is at the shot-noise limit, $\delta\hat{s}_{in}$ has the same noise correlators as $\delta\hat{s}_{vac}$. In such situations, the total cavity noise is said to be due to photon shot noise (with contribution from both the input optical mode and the dissipation channels link to the external bath). The output optical field is given by \cite{Marquardt_Arxiv_2013}
\begin{equation}\label{eq:sout}
\hat{s}_{out}=\hat{s}_{in}-\sqrt{2\kappa_e}\hat{a} ~.
\end{equation}
In the Fabry-P\'{e}rot cavity setting, the output mode is the reflected light from the cavity (see Fig. \ref{fig:setup}~(a)), while in integrated micro-resonators, it is the waveguide mode at the thru-port (see Fig. \ref{fig:setup}~(b)). We denote the intra-cavity photon number by $\hat{n}=\hat{a}^\dagger\hat{a}$, and the photon flow rate at the input and output by $\hat{P}_{in}=\hat{s}_{in}^\dagger\hat{s}_{in}$ and $\hat{P}_{out}=\hat{s}_{out}^\dagger\hat{s}_{out}$.

Eq. \eqref{eq:QLE_mech} describes the dynamics of the mechanical resonator. $\gamma$ is the mechanical damping rate defined as the HWHM of the resonator response. It is related to the mechanical quality factor $Q_M$ by $Q_M=\Omega_M/2\gamma$. The term $g\hat{a}^\dagger\hat{a}$ represents the optical force acting on the mechanical resonator. $\delta\hat{f}_{th}$ is the thermomechanical fluctuation force 
at thermal equilibrium and has a correlation function of \cite{Vitali_PRA_2001}
\begin{equation}\label{eq:noise_corr_thermal}
\begin{aligned}
&\langle\delta\hat{f}_{th}(t) \delta\hat{f}_{th}(t^\prime)\rangle \\
&~~~~= \int{\frac{d\omega}{2\pi} e^{-i\omega(t-t^\prime)} \frac{\omega}{\Omega_M} \left[\coth\left(\frac{\hbar\omega}{2k_B T}\right)+1\right]} ~.
\end{aligned}
\end{equation}
For a resonator with high $Q_M$, $\delta\hat{f}_{th}(t)$ can be approximated as $\delta$-correlated \cite{Kac_PRL_1981}, i.e., $\langle\delta\hat{f}_{th}(t) \delta\hat{f}_{th}(t^\prime) +\delta\hat{f}_{th}(t^\prime) \delta\hat{f}_{th}(t)\rangle \approx 2(2\bar{n}_{th}+1)\delta(t-t^\prime)$, where $\bar{n}_{th} = (e^{\hbar\Omega_M/k_B T}-1)^{-1} \approx k_B T/\hbar\Omega_M$ is the average thermal phonon occupation.

\begin{figure}[t]
\centering
\includegraphics[width=8cm]{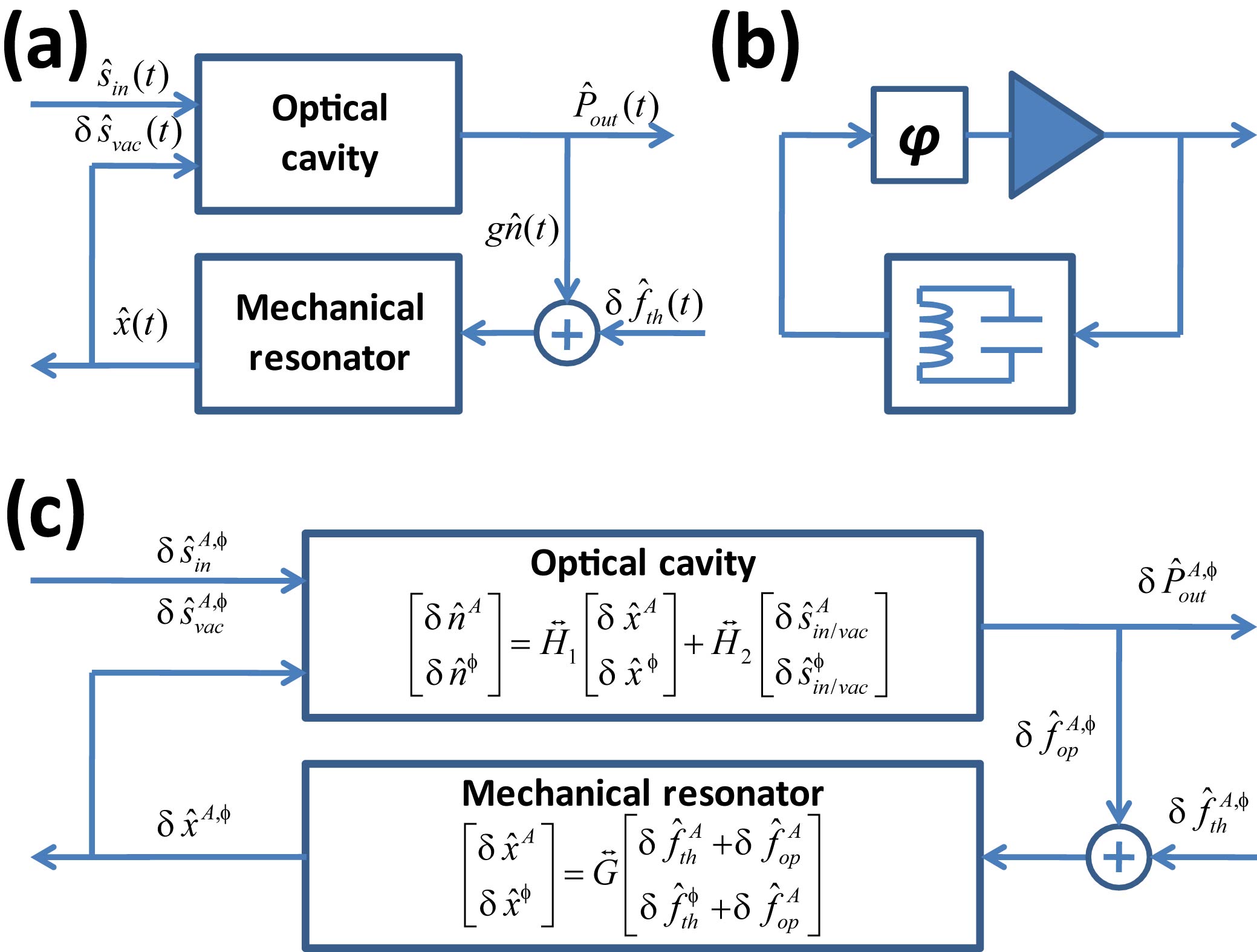}
\caption{ (a) Feedback loop of a cavity optomechanical system. (b) Feedback loop of a typical electrical oscillator. (c) Feedback loop for the small-signals in a cavity optomechanical system. }
\label{fig:feedback_loop}
\end{figure}

The two coupled quantum Langevin equations in Eqs. \eqref{eq:QLE_opt} and \eqref{eq:QLE_mech} can be viewed as representing two components that form a feedback loop \cite{Stamper-Kurn_PRA_2012, vanderZant_PhysRep_2012}. While the mechanical motion modulates the cavity field by changing the optical detuning (as described by Eq. \eqref{eq:QLE_opt}), the modulation of the cavity field generates an optical force which feeds back to drive the mechanical motion (as described by Eq. \eqref{eq:QLE_mech}). This feedback loop is illustrated by the block diagram in Fig. \ref{fig:feedback_loop}~(a). The inputs of the feedback loop include the laser input $s_{in}$, vacuum fluctuation $\delta\hat{s}_{vac}$, and the thermomechanical fluctuation force $\delta\hat{f}_{th}$. The outputs of the loop are the displacement $\hat{x}$, the intra-cavity photon number $\hat{n}$ and the output photon flow $\hat{P}_{out}$. $\hbar\Omega_L\hat{P}_{out}$ is the output optical power that can be measured directly in experiment. In principle, the output field $\hat{s}_{out}$ can also be detected using the optical homodyne method which mixes the output light with a strong local oscillator. This method has been applied in experiments to monitor the mechanical motion with high sensitivity, see for example \cite{Kippenberg_Nature_2012}. In oscillator applications, a direct measurement of $\hbar\Omega_L\hat{P}_{out}$ is more commonly used (see for example \cite{Bhave_OE_2011, Tang_APL_2012, Lin_OE_2012}) because it does not require an extra optical path for the local oscillator and so facilitates a more compact device design, which is more desirable for integrated purposes. In this paper we will therefore mainly consider $\hat{P}_{out}$ as the oscillator output but the results can be easily generalized to describe the situations using other detection schemes.

The feedback loop of an optomechanical system has a loop gain that can be positive or negative, depending on the detuning of the cavity. As have been well studied, a blue (red) detuned cavity provides a positive (negative) feedback which effectively heats up (cools down) the mechanical motion. For the situation of positive feedback, the system becomes unstable when the loop gain is larger than unity. In that case even a small fluctuation is largely amplified and the oscillating amplitude grows until it reaches a limit-cycle set by the nonlinearity of the system. Note that the two equations of motion Eqs.~\eqref{eq:QLE_opt} and \eqref{eq:QLE_mech} are nonlinear and therefore the limiting mechanism is intrinsically included \cite{Tang_PRA_2012}. The optical cavity and mechanical resonator resemble the two essential components of an oscillator loop: the optical cavity acts as an amplifier which provides optical force to amplify the mechanical motion and the mechanical resonator acts as a frequency selective element which selects out only a narrow frequency band. For comparison, Fig. \ref{fig:feedback_loop}~(b) shows a typical electronic oscillator loop. For characterization of an oscillator, phase noise is the foremost important figure of merit \cite{Rubiola_Book_2010}. In this paper, we focus on the phase noise behavior of such self-sustained optomechanical oscillation.

\subsection{Small-signal analysis and spectral filter decomposition}\label{subsec:smallsignal}

A standard approach to solve for the noise response of a feedback loop is to use transfer function method \cite{Rubiola_Book_2010}. After solving for the transfer function for each of the loop components, they can be combined to describe the closed-loop response. However, since the components forming the feedback loop of an optomechanical system are nonlinear, in order to make use of the traditional linear transfer function method, the equations of motion need to be first linearized. In the vast majority of theoretical analyses of optomechanical systems, the equations of motion are linearized around a static solution (assuming $\hat{x}(t)=x_0+\delta\hat{x}(t)$) and the oscillating signal $\delta\hat{x}$ is treated as small perturbation. However in the case of self-sustained oscillation, the oscillating amplitude can no longer be assumed to be small. Instead, here we linearize the equations around a limit-cycle solution.

\begin{figure}[t]
\centering
\includegraphics[width=8.6cm]{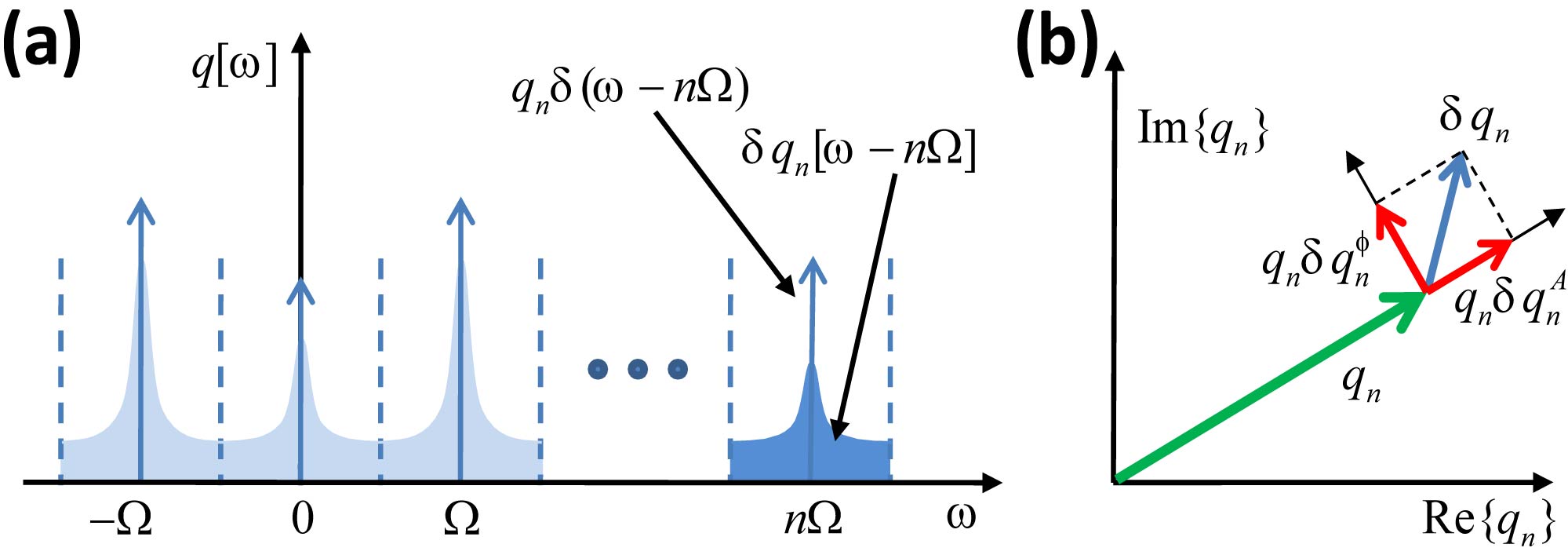}
\caption{ (a) Illustration of the spectral-filter decomposition described in Eq. \eqref{eq:dq_sum}. (b) Illustration of the amplitude and phase quadrature described in Eq. \eqref{eq:dq_A_phi}. }
\label{fig:smallsignal_analysis}
\end{figure}

Let us assume the system has reached a limit-cycle and is oscillating at frequency $\Omega$. We assume $\Omega$ is close to the mechanical resonance frequency, i.e., $|\Omega-\Omega_m|\ll\Omega_m$ and we limit our analysis only to the situations where there is no period-doubling or chaotic behavior, such as the situations described in Ref. \cite{Vahala_PRL_2007}. We also assume the linewidth of the mechanical resonator is smaller than that of the optical cavity, i.e., $\gamma<\kappa$, which is valid for most practical situations. At the limit-cycle, the average dynamics of the quantities such as $\hat{x}(t)$, $\hat{a}(t)$, $\hat{n}(t)$, $\hat{s}_{out}(t)$, and $\hat{P}_{out}(t)$ follow a periodic path with frequency $\Omega$. Let us use $\hat{q}(t)$ to represent any of these quantities and write it as a sum of a periodic function and a small-signal, i.e., $\hat{q}(t)=q_{cyc}(t)+\delta\hat{q}(t)$. The limit-cycle solution $q_{cyc}(t)$ describes the classical average dynamics $q_{cyc}(t)=\langle\hat{q}(t)\rangle$ and is expressed as a complex scalar function. The small-signal $\delta\hat{q}(t)$ is an operator which preserves the quantum description and has a zero expectation value $\langle\delta\hat{q}(t)\rangle=0$. Throughout the manuscript, we will interchangeably refer $q_{cyc}(t)$ as ``limit-cycle solution'' or ``coherent amplitude'', and refer $\delta\hat{q}(t)$ as ``small-signal'', ``noise'', or ``perturbation''. Since $q_{cyc}(t)$ is periodic in $\Omega$, it can be expressed in Fourier series as
\begin{equation}\label{eq:q_cyc}
q_{cyc}(t)=\sum_n e^{-in\Omega t} q_n ~.
\end{equation}
In the notation $\sum_n\equiv\sum_{n=-\infty}^{+\infty}$ the sum runs from negative infinity to positive infinity if not otherwise specified. Note that $q_{cyc}(t)$ is periodic but not necessarily sinusoidal, i.e., it may have nonzero higher harmonic components. In practice, all oscillator systems work in the nonlinear regime and therefore may have non-negligible higher harmonic components. In fact, as have been shown in Ref. \cite{Tang_PRA_2012}, for an optomechanical oscillator operating at large amplitude regime, the higher harmonic components of the cavity field play an important role in the system dynamics. As we will see later, they are also crucial in determining the noise performance. Therefore in the present analysis keeping all the harmonic terms in Eq. \eqref{eq:q_cyc} is necessary. For the small-signal term $\delta\hat{q}(t)$, one can also write it in a Fourier series-like expansion as
\begin{equation}\label{eq:dq_n}
\delta\hat{q}(t)=\sum_n e^{-in\Omega t} \delta\hat{q}_n(t) ~.
\end{equation}
However, since $\delta\hat{q}(t)$ is in general not periodic, the harmonic components $\delta\hat{q}_n(t)$ have a time dependence and they are not uniquely defined. Here, we define $\delta\hat{q}_n(t)=\frac{\Omega}{2\pi}\mathrm{sinc}\left(\frac{\Omega t}{2}\right)\otimes[\delta\hat{q}(t)e^{in\Omega t}]$,
where $\otimes$ denotes the convolution operation. In frequency domain, it can be expressed as $\delta\hat{q}_n[\omega]=\delta\hat{q}[\omega+n\Omega] \mathrm{rect}(\omega/\Omega)$, where $\mathrm{rect}(x)$ is the rectangular function which has a value of one when $-1/2<x\le1/2$ and zero otherwise. Throughout the manuscript, a square bracket notation $f[\omega]$ is used to represent the Fourier transform of the time domain signal $f(t)$, i.e., $f[\omega]=\mathcal{F}\{f(t)\}$. From the frequency domain expression, it is apparent that $\delta\hat{q}_n(t)$ has a frequency spectrum which is non-zero only within the range $-\Omega/2<\omega\le\Omega/2$. In engineering terminology, it is a baseband function with a bandwidth of $\Omega$. We call this operation described by Eq. \eqref{eq:dq_n} the ``spectral filter decomposition''. Details of the spectral filter decomposition and its mathematical implications can be found in Appendix~\ref{app:spectral_decomposition}. As we will see later, such a baseband definition is important as it ensures that each of the $\delta\hat{q}_n(t)$ is varying at a time scale slower than the oscillation frequency $\Omega$ and it guarantees that there is no spectral overlapping when the components are mixed by nonlinearities. Also, under such definition it can be shown that the cross power spectral densities between different harmonic components are given by the following simple form,
\begin{equation}\label{eq:dq_n_psd}
\begin{aligned}
\mathcal{S}_{\delta\hat{q}_n \delta\hat{q}_m}[\omega] &= \mathcal{S}_{\delta\hat{q}\delta\hat{q}}[\omega+n\Omega] \mathrm{rect}[\omega/\Omega] \delta_{n,-m} \\
\mathcal{S}_{\delta\hat{q}_n \delta\hat{q}_m^\dagger}[\omega] &= \mathcal{S}_{\delta\hat{q}\delta\hat{q}^\dagger}[\omega+n\Omega] \mathrm{rect}[\omega/\Omega] \delta_{n,m} \\
\mathcal{S}_{\delta\hat{q}_n^\dagger \delta\hat{q}_m}[\omega] &= \mathcal{S}_{\delta\hat{q}^\dagger\delta\hat{q}}[\omega-n\Omega] \mathrm{rect}[\omega/\Omega] \delta_{n,m} \\
\mathcal{S}_{\delta\hat{q}_n^\dagger \delta\hat{q}_m^\dagger}[\omega] &= \mathcal{S}_{\delta\hat{q}^\dagger\delta\hat{q}^\dagger}[\omega-n\Omega] \mathrm{rect}[\omega/\Omega] \delta_{n,-m} ~,
\end{aligned}
\end{equation}
where the cross power spectral densities are defined as $\mathcal{S}_{\hat{X}\hat{Y}}[\omega]=\int d\tau e^{i\omega\tau}\langle\hat{X}(t+\tau)\hat{Y}(t)\rangle$. As expected, harmonic components from different frequency range are uncorrelated. 
In summary, we can write
\begin{equation}\label{eq:dq_sum}
\hat{q}(t)=\sum_n e^{-in\Omega t} \left[q_n+\delta\hat{q}_n(t)\right] ~.
\end{equation}
The result of expanding a function in the form of Eq. \eqref{eq:dq_sum} is illustrated in Fig. \ref{fig:smallsignal_analysis}~(a). It can be understood as a sum of harmonics each oscillating at its own frequency and having a baseband noise spectrum associated with it. With the assumption $|q_n|^2\gg\langle\delta\hat{q}_n^2\rangle$, we define
\begin{equation}\label{eq:dq_A_phi}
\begin{aligned}
\delta\hat{q}_n^A(t) &=& \frac{1}{2} \left[ \frac{\delta\hat{q}_n(t)}{q_n} +\frac{\delta\hat{q}_n^\dagger(t)}{q_n^\ast} \right] \\
\delta\hat{q}_n^\phi(t) &=& \frac{1}{2i} \left[ \frac{\delta\hat{q}_n(t)}{q_n} -\frac{\delta\hat{q}_n^\dagger(t)}{q_n^\ast}\right] ~.
\end{aligned}
\end{equation}
By definition these two operators are Hermitian. Using these two quantities to rewrite Eq. \eqref{eq:dq_sum} as $\hat{q}(t)=\sum_n e^{-in\Omega t+i\delta\hat q_n^\phi(t)}q_n[1+\delta\hat q_n^A(t)]$ (keeping only first order terms), it becomes apparent that $\delta\hat q_n^A(t)$ and $\delta\hat q_n^\phi(t)$ can be identified as the relative amplitude noise and phase noise with respect to the carrier signal $e^{-in\Omega t}q_n$. Fig. \ref{fig:smallsignal_analysis}~(b) illustrates the meaning of the definitions of Eq. \eqref{eq:dq_A_phi}. We can see that $\delta\hat q_n^A(t)$ and $\delta\hat q_n^\phi(t)$ are the in-phase and quadrature components rotated and normalized with respect to $q_n$.

The above analysis applies to any periodic function of time. (The spectral filter decomposition (Eqs.~\eqref{eq:dq_n} and \eqref{eq:dq_n_psd} applies even when the function is not periodic.) In the next section we apply the above analysis to the quantities $\hat{b}(t)$, $\hat{a}(t)$, $\hat{n}(t)$, $\hat{s}_{out}(t)$, and $\hat{P}_{out}(t)$, and derive the transfer functions relating their relative amplitude noise and phase noise. We emphasize that since now we are carrying out perturbation analysis around a periodic solution, for each dynamical quantity there are two independent perturbation directions, namely the in-phase and quadrature directions (or amplitude and phase directions). Therefore, transfer functions relating any two perturbation terms are $2\times2$ matrices taking into account all the inter-quadratures transfers. This concept is illustrated in the feedback loop for the linearized system in Fig. \ref{fig:feedback_loop}~(c). The traditional approach to analyze oscillator phase noise is to apply the transfer function method in phase noise space and ignore the effect of amplitude noise \cite{Rubiola_Book_2010}. Here by using transfer matrices we include the contribution from amplitude noise and all the amplitude-to-phase inter-transfers.

Before we derive the transfer functions, there are two remarks to be addressed. First, for the mechanical displacement, since we are considering only the situations where the oscillating frequency is close to the mechanical resonance frequency, i.e., $|\Omega-\Omega_m|\ll\Omega_m$, as it is the case for most practical optomechanical oscillator systems, 
it can be shown that effect of forces acting on the resonator at high harmonic frequency $n\Omega$ are reduced by a factor of $(n^2-1)Q_M$ relative to the first harmonic. Therefore, for a system with reasonably high $Q_M$, we can safely neglect the higher harmonic terms. We also take the static displacement $x_0$ to be zero, since its effect is equivalent to a static shift in cavity detuning. Therefore, we keep only the first harmonic terms for $\hat{x}(t)$ and the expression becomes
\begin{equation}\label{eq:x1}
\hat{x}(t)=e^{-i\Omega t} [x_1+\delta\hat{x}_1(t)] + h.c. ~,
\end{equation}
$h.c.$ denotes the Hermitian conjugates of all previous terms in the equation. 

Second, 
the laser noise $\delta\hat{s}_{in}$ and the vacuum fluctuation $\delta\hat{s}_{vac}$ can be expressed in harmonic components using the spectral filter decomposition described in Eq. \eqref{eq:dq_n} as
\begin{equation}\label{eq:def:dq_n}
\delta\hat{s}_{in/vac}(t)=\sum_n e^{-in\Omega t} \delta\hat{s}_{in/vac,n}(t) ~.
\end{equation}
According to Eqs. \eqref{eq:noise_corr_vac} and \eqref{eq:dq_n_psd}, for the cavity vacuum fluctuation and the laser shot noise the only nonzero correlator is given by $\mathcal{S}_{\delta\hat{s}_{vac,n} \delta\hat{s}_{vac,m}^\dagger}[\omega] = \mathrm{rect}[\omega/\Omega] \delta_{n,m}$. The relative amplitude noise and phase noise of the input laser as different spectral range is defined as
\begin{equation}\label{eq:laser_noise}
\begin{aligned}
\delta\hat{s}_{in,n}^A(t) &= \frac{1}{2} \left[ \frac{\delta\hat{s}_{in,n}(t)}{s_{in,0}} +\frac{\delta\hat{s}_{in,n}^\dagger(t)}{s_{in,0}^\ast} \right] \\
\delta\hat{s}_{in,n}^\phi(t) &= \frac{1}{2i} \left[ \frac{\delta\hat{s}_{in,n}(t)}{s_{in,0}} -\frac{\delta\hat{s}_{in,n}^\dagger(t)}{s_{in,0}^\ast}\right] ~.
\end{aligned}
\end{equation}
The difference between Eqs.~\eqref{eq:dq_A_phi} and \eqref{eq:laser_noise} is that the laser noise is normalized and rotated with respect to the laser field $s_{in}$ instead of the individual harmonic components.

\begin{figure}[t]
\centering
\includegraphics[width=8cm]{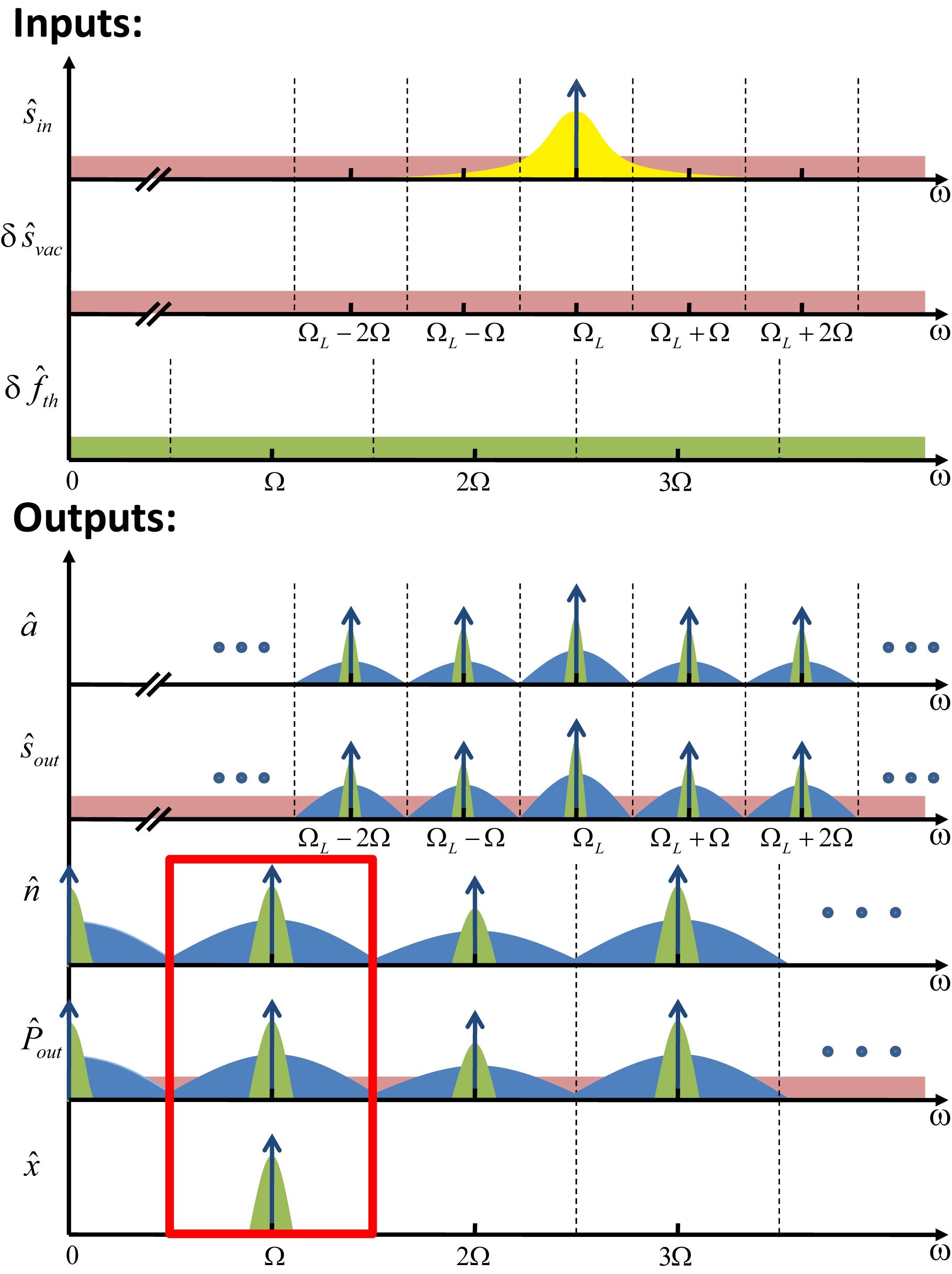}
\caption{ A qualitative illustration of the frequency spectra of all the relevant input and output quantities. The arrows represent the delta-function spectrum of the coherent amplitude and the shaded areas represent the spectral densities of the noise. Dashed lines are used to separate different regions under the spectral filter decomposition. The quantities $\hat{s}_{in}$, $\delta\hat{s}_{vac}$, $\hat{a}$, and $\hat{s}_{out}$ are expressed in the non-rotating frame to emphasize the original frequency scale. The three spectra of interest for characterizing the oscillator performance are highlighted by the red bounding box ($\hat{n}_1[\omega]$, $\hat{P}_{out,1}[\omega]$, and $\hat{x}_1[\omega]$).  }
\label{fig:spectrum}
\end{figure}

To summarize the analysis in this section, we illustrate qualitatively in Fig. \ref{fig:spectrum} the expected frequency spectra of all the input/output quantities. In the figure, we use arrows to represent the delta-function spectrum of the coherent amplitude ($q_n$) and use shaded areas to represent the spectral densities of the noise ($\delta\hat{q}_n$). We rotate $\hat{a}$, $\hat{s}_{in}$, and $\hat{s}_{out}$ back to the non-rotating frame to emphasize the original frequency scale. Dashed-lines indicate the regions under the spectral filter decomposition. For the inputs, the laser input $\hat{s}_{in}$ has one single coherent amplitude representing the input field $s_{in}$. Its noise spectrum consists of a white shot noise background and an additional technical laser noise. In practice all lasers have low-frequency noise with a $1/f^n$ dependent spectrum featuring noise with long correlation time. This technical laser noise is ``low-frequency'' compared to the laser frequency but could be comparable or larger than the mechanical frequency $\Omega$. In particular, semiconductor diode lasers can exhibit phase noise up to 30 dB above the shot noise level at GHz frequencies due to the electrons relaxation dynamics \cite{Gorodetsky_NJP_2013}. For simplicity, in this manuscript we only consider low-frequency laser noise that is non-negligible within the range $|\omega|<\Omega/2$, which is practically achievable by filtering the laser noise with narrowband filters. Beside the laser noise, vacuum noise enters the system through the cavity intrinsic dissipation channel. It has a white spectrum and has the same value as the laser shot noise spectrum. Another noise input are the thermomechanical fluctuations $\delta\hat{f}_{th}$, which are also white and have a cutoff frequency much larger than the mechanical frequency.

On the outputs side, the coherent amplitude of the intra-cavity field $\hat{a}$ and the output optical field $\hat{s}_{out}$ contain many harmonics due to scattering by the mechanical oscillation, as has been discussed in Ref. \cite{Tang_PRA_2012}. For each harmonic peak, there is a noise spectrum associated with it. For $\hat{a}$, one may expect that the noise spectrum shows two characteristic frequency scales: when the frequency offset from the carrier $|\omega-n\Omega|$ is within $\gamma$ the noise shows signature of the mechanical response (green regions in Fig. \ref{fig:spectrum}) and when the frequency offset from the carrier is beyond $\kappa$ the noise decays since the cavity acts like a bandpass filter which has lower response outside the bandwidth (blue regions in Fig. \ref{fig:spectrum}). In the figure the system is assumed to be in the resolved sideband regime (RSR), $\kappa<\Omega$. In the unresolved sideband regime (USR), $\kappa>\Omega$, one can imagine that there is significant overlap between the blue regions. The spectrum of the output field $\hat{s}_{out}$ is similar but has an additional broadband background (red regions in Fig. \ref{fig:spectrum}) since according to Eq. \eqref{eq:sout} the laser noise can directly go to the output field. The cavity photon number $\hat{n}=\hat{a}^\dagger\hat{a}$ and the output photon flow $\hat{P}_{out}=\hat{s}_{out}^\dagger\hat{s}_{out}$ are directly related to $\hat{a}$ and $\hat{s}_{out}$ and so their spectra display similar features but occur at the mechanical frequency instead of the optical frequency. Lastly, for the mechanical displacement $\hat{x}$, the mechanical response decays quickly at high harmonic frequencies so it only has significant response around $\Omega\approx\Omega_M$.

When one talks about phase noise of a self-sustained optomechanical osillator, one refers to the phase noise spectrum of the measured output power $\hbar\Omega_L\hat{P}_{out}$ at the first harmonic frequency $\Omega$. Besides this, we are also interested in the phase noise spectrum of $\hat{x}_1$ and $\hat{n}_1$ at the first harmonic since they represent the displacement and optical force of the mechanical resonator. These three noise spectra highlighted by a red bounding box in Fig. \ref{fig:spectrum} are the spectra of interest for characterizing the performance of an optomechanical oscillator. In the following we will first derive the transfer functions for all the output quantities and then study in details the behavior of these three noise spectra.

\subsection{Derivation of the small-signal transfer functions}\label{subsec:tfunc}

From the equations of motion (Eqs. \ref{eq:QLE_opt} and \ref{eq:QLE_mech}), first we observe that they are autonomous, i.e., without explicit time dependence, which implies that the system is time translational invariant. This trivial observation has a non-trivial consequence in self-oscillating systems. In an oscillating system, a shift in time by $t\rightarrow t+\Delta t$ is equivalent to a shift in phase by $\theta\rightarrow\theta+n\Omega\Delta t$ in the $n$th-harmonics of all the dynamical quantities. This phase translational invariance implies that the phase can diffuse without any restoring constraint. As we will see later it causes a $1/\omega^2$ dependence in the phase noise spectrum. It also means that there is an arbitrary choice of reference phase. Without loss of generality, we choose the phase of $x_1$ to be zero, which means the resonator has maximum displacement at $t=0$.

Let us first look at the equation of motion for the optical cavity (Eq. \eqref{eq:QLE_opt}). By substituting $\hat{a}=a_{cyc}+\delta\hat{a}$, $\hat{s}_{in}=s_{in}+\delta\hat{s}$, and Eq. \eqref{eq:x1} for $\hat{x}(t)$ and keeping the perturbation terms up to first order, we have
\begin{align}\label{eq:ODE_a_cyc}
&\dot{a}_{cyc}=[-\kappa+i\Delta+i2gx_1\cos\Omega t] a_{cyc} +\sqrt{2\kappa_e}s_{in} \\
&\begin{aligned}\label{eq:ODE_da}
\delta\dot{\hat{a}}&=[-\kappa+i\Delta+i2gx_1\cos\Omega t]\delta\hat{a} +\sqrt{2\kappa_e}\delta\hat{s}_{in} \\
&+\sqrt{2\kappa_i}\delta\hat{s}_{vac} +ig(e^{-i\Omega t}\delta\hat{x}_1 + e^{i\Omega t}\delta\hat{x}_1^\dagger)a_{cyc} ~.
\end{aligned}
\end{align}
These two first-order ordinary differential equations can be solved by the standard integrating factor method. After the solution of $a_{cyc}$ is found, it can be substituted into Eq. \eqref{eq:ODE_da} to solve for $\delta\hat{a}$. It can be shown that the solutions of 
$a_n$ and $\delta\hat{a}_n$ in the frequency domain are given by
\begin{subequations}\label{eq:a_tfunc}
\begin{align}
\label{eq:a_n}
&a_n=\sqrt{2\kappa_e}s_{in}\sum_m\frac{J_m J_{m-n}}{K_m} \\
\label{eq:da_n}
&\delta\hat{a}_n[\omega] =\sum_{m,p} \frac{J_{m-n}J_{m-p}}{K_m} L_{K_m}[\omega] \sqrt{2\kappa_e}\delta\hat{s}_{in,p}[\omega] \nonumber\\
&+\sum_{m,p} \frac{J_{m-n}J_{m-p}}{K_m} L_{K_m}[\omega] \sqrt{2\kappa_i}\delta\hat{s}_{vac,p}[\omega] \\
&+\sqrt{2\kappa_e}s_{in}\sum_m{\frac{igJ_{m-n}J_{m-1}}{K_m K_{m-1}}} L_{K_m}[\omega] \delta\hat{x}_1[\omega] \nonumber\\
&+\sqrt{2\kappa_e}s_{in}\sum_m{\frac{igJ_{m-n}J_{m+1}}{K_m K_{m+1}}} L_{K_m}[\omega] \delta\hat{x}_1^\ddagger[\omega] ~, \nonumber
\end{align}
\end{subequations}
where $J_n\equiv J_n(\tilde{x})$ is the $n$-th order Bessel function of the first kind with the argument omitted for brevity, the dimensionless amplitude $\tilde{x}$ is defined as $\tilde{x}=2g|x_1|/\Omega$, $K_m$ is a complex quantity defined as
\begin{equation}
K_m=\kappa-i(\Delta+m\Omega) ~,
\end{equation}
and $L_z[\omega]$ is the normalized first-order filter function defined as
\begin{equation}
L_z[\omega]=\frac{z}{z-i\omega} ~.
\end{equation}
The absolute square of $L_z[\omega]$ is a Lorentzian function with the linewidth and the center frequency given by $\mathrm{Re}\{z\}$ and $\mathrm{Im}\{z\}$ and is one at $\omega=0$. Here, we use the double dagger notation to denote the Fourier transform of the Hermitian conjugate, i.e., $\delta\hat{x}_1^\ddagger[\omega]=(\delta\hat{x}_1[-\omega])^\dagger =\mathcal{F}\{\delta\hat{x}_1^\dagger(t)\}$. In the derivation, the Jacobi-Anger expansion $e^{iz\sin\theta}=\sum_n J_n(z)e^{in\theta}$ and the Bessel function identity $\sum_n J_n J_{n+m}=\delta_{m,0}$ were used.

\begin{figure}[t]
\centering
\includegraphics[width=8.6cm]{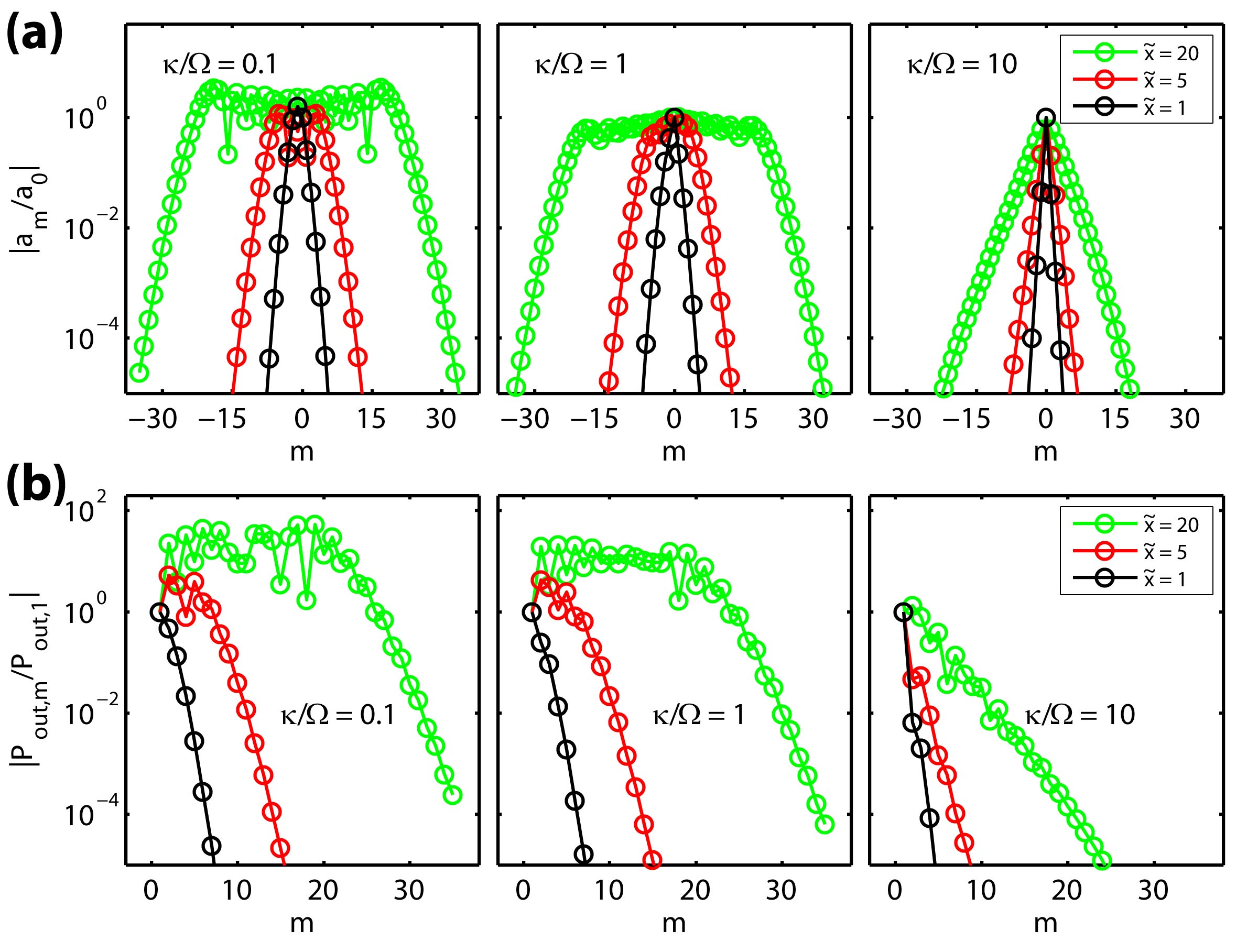}
\caption{ Relative magnitude of the higher harmonic components of (a) $a_m$, (b) $n_m$, and (c) $P_{out,m}$ plotted against index $m$ at different $\kappa/\Omega$ and $\tilde{x}$. In all the calculations, the optimal detuning described by the white dashed line in Fig. \ref{fig:LCamp} (a) is used. For the calculation of $P_{out,m}$, critical coupling condition $\kappa_e=\kappa_i$ is assumed. }
\label{fig:AnNnPn}
\end{figure}

Eq. (\ref{eq:a_n}) was previously derived in Refs. \cite{Girvin_PRL_2006} and \cite{Tang_PRA_2012}. It describes how the harmonics of the intra-cavity field are related to other parameters of the system. In particular, it has a dependence on the displacement amplitude $\tilde{x}$ through the Bessel functions $J_n(\tilde{x})$. When the displacement amplitude is small, i.e., $\tilde{x}\ll n$, the Bessel function can be approximated as $J_n(\tilde{x})\approx(\tilde{x}/2)^n/n!$ for positive index $n$. In such case, it can be shown that the magnitude of $a_n$ is of the order of $\tilde{x}^n$. Therefore in the small amplitude limit, only the low harmonic components remain. When $\tilde{x}\gtrsim n$ the above approximation is no longer valid and the higher harmonic components could have significantly large magnitudes. Therefore the value of $\tilde{x}$ is a good indicator of the onset of the optical nonlinearity due to large displacement. To give an idea on the magnitude of the higher harmonic components, we plot $|a_n/a_0|$ for different $\kappa/\Omega$ and $\tilde{x}$ in Fig. \ref{fig:AnNnPn}~(a).

Because of the presence of the higher harmonic components, the laser noise and the vacuum noise can be scattered to different frequency ranges and mixed with each other, and that is exactly what is described by Eq. \eqref{eq:da_n}. We can see that the noise $\delta\hat{a}_n$ at $n$-th harmonics has contribution from $\delta\hat{s}_{in/vac,p}$ in all other harmonics, as well as the first harmonic displacement noise $\delta\hat{x}_1$. Note that the noise transfers always go through the filter functions $L_{K_m}[\omega]$, which reflects the fact that the optical cavity acts like a filter filtering out the noises outside the cavity linewidth $\kappa$.

We mentioned in previous section that the noise transfer function can be written in a $2\times2$ matrix equation. To see how it can be done, we point out that one can obtain an expression for $\delta\hat{a}_n^\ddagger$ by taking Hermitian conjugate on both sides of the Eq. \eqref{eq:da_n}. Therefore, Eq. \eqref{eq:da_n} can be viewed as two equations: one for $\delta\hat{a}_n$ and one for $\delta\hat{a}_n^\ddagger$. It is analogous to the fact that a complex number equation represents one equation for the real part and another equation for the imaginary part. Instead of writing the two equations in terms of $\delta\hat{a}_n$ and $\delta\hat{a}_n^\ddagger$, one can also use $\delta\hat{a}^A$ and $\delta\hat{a}^\phi$ defined in Eq. \eqref{eq:dq_A_phi} as the two independent quantities. In either way, Eq. \eqref{eq:da_n} can be rewritten as a $2\times2$ matrix equation, as illustrated in Fig. \ref{fig:feedback_loop}~(c).

After the cavity field $\hat{a}$ is solved, the output field $\hat{s}_{out}$, the photon number $\hat{n}$ and the output photon flow $\hat{P}_{out}$ can be derived from it. The expression for $\hat{s}_{out}$ can be derived using Eq. \eqref{eq:sout} and is given by
\begin{subequations}\label{eq:sout_tfunc}
\begin{align}
\label{eq:sout_n}
s_{out,n} &= \delta_{n,0}s_{in} -\sqrt{2\kappa_e}a_n \\
\label{eq:dsout_n}
\delta\hat{s}_{out,n}[\omega] &= \delta\hat{s}_{in,n}[\omega] -\sqrt{2\kappa_e}\delta\hat{a}_n[\omega] ~.
\end{align}
\end{subequations}
For the cavity photon number $\hat{n}=\hat{a}^\dagger\hat{a}$, we have $n_n=\sum_m{a_m^\ast a_{m+n}}$ and $\delta\hat{n}_n=\sum_m(a_m^\ast\delta\hat{a}_{m+n}+a_{m+n}\delta\hat{a}_m^\dagger)$, and therefore,
\begin{subequations}\label{eq:n_tfunc}
\begin{align}
\label{eq:n_n}
&n_n = 2\kappa_e P_{in}\sum_m\frac{J_m J_{m+n}}{K_m^\ast K_{m+n}} \\
\label{eq:dn_n}
&\delta\hat{n}_n[\omega] = 2\kappa_e P_{in}\times \nonumber\\
& \left\{\sum_{m,p}{\frac{J_{m-n}J_{m-p} L_{K_m}}{K_{m-n}^\ast K_m}} \left[\frac{\delta\hat{s}_{in,p}}{s_{in}}+\sqrt{\frac{\kappa_i}{\kappa_e}} \frac{\delta\hat{s}_{vac,p}}{s_{in}}\right]\right. \nonumber\\
&+ \sum_{m,p} \frac{J_{m+n}J_{m-p}L_{K_m^\ast}}{K_{m+n}K_m^\ast} \left[\frac{\delta\hat{s}_{in,p}^\ddagger}{s_{in}^\ast}+\sqrt{\frac{\kappa_i}{\kappa_e}} \frac{\delta\hat{s}_{vac,p}^\ddagger}{s_{in}^\ast}\right] \\
&+ \sum_m
\frac{ig J_m J_{m+n-1}}{K_m^\ast K_{m+n-1}} \left[\frac{L_{K_{m+n}}}{K_{m+n}}-\frac{L_{K_{m-1}^\ast}}{K_{m-1}^\ast}\right] \delta\hat{x}_1 \nonumber\\
&\left.+ \sum_m
\frac{ig J_{m}J_{m+n+1}}{K_m^\ast K_{m+n+1}} \left[\frac{L_{K_{m+n}}}{K_{m+n}}-\frac{L_{K_{m+1}^\ast}}{K_{m+1}^\ast}\right] \delta\hat{x}_1^\ddagger\right\} ~. \nonumber
\end{align}
\end{subequations}
In the above expression we have omitted the frequency dependence for simplicity. To solve for $\hat{P}_{out}$, we make use of the equation $\hat{P}_{out} =-\dot{\hat{n}} -2\kappa_i\hat{n} +\hat{P}_{in} +\sqrt{2\kappa_i} (\hat{a}^\dagger\delta\hat{s}_{vac} +\delta\hat{s}_{vac}^\dagger\hat{a})$, which can be verified by combining Eqs. \eqref{eq:QLE_opt} and \eqref{eq:sout}. It follows that the expressions for $\hat{P}_{out}$ are given by
\begin{subequations}\label{eq:Pout_tfunc}
\begin{align}
\label{eq:Pout_n}
&P_{out,n}=(in\Omega-2\kappa_i) n_n + P_{in}\delta_{n,0} \\
&\begin{aligned}\label{eq:dPout_n}
&\delta\hat{P}_{out,n}[\omega] =(in\Omega+i\omega-2\kappa_i) \delta\hat{n}_n[\omega] \\
&+(s_{in}^\ast\delta\hat{s}_{in,n}[\omega] +s_{in}\delta\hat{s}_{in,-n}^\ddagger[\omega]) \\
&+\sqrt{2\kappa_i} \sum_m(a_{m}^\ast\delta\hat{s}_{vac,n+m}[\omega] +a_{n+m}\delta\hat{s}_{vac,m}^\ddagger[\omega]) ~.
\end{aligned}
\end{align}
\end{subequations}
Fig. \ref{fig:AnNnPn} (b) shows $|P_{out,n}/P_{out,1}|$ for different $\kappa/\Omega$ and $\tilde{x}$. In the plots, critical coupling condition $\kappa_e=\kappa_i$ is assumed. Next, let us consider the equation of motion for the mechanical resonator (Eq. \eqref{eq:QLE_mech}). As discussed before, we only need to consider the first harmonic component of the displacement and so the optical force. Since $|\Omega-\Omega_M|\ll\Omega_M$ and $\gamma\ll\Omega_M$ is assumed, it can be shown that
\begin{subequations}\label{eq:x_tfunc}
\begin{align}
\label{eq:x_1}
x_1&=\frac{1}{\Gamma}ign_1\\
\label{eq:dx_1}
\delta\hat{x}_1[\omega]&=\frac{iL_\Gamma[\omega]}{\Gamma}\left(g \delta\hat{n}_1[\omega]+\sqrt{2\gamma}\delta\hat{f}_{th,1}[\omega]\right) ~,
\end{align}
\end{subequations}
where $\Gamma=\gamma-i\nu$ and $\nu=\Omega-\Omega_M$ is the mechanical detuning. This time the noise transfer function is a filter function $L_\Gamma[\omega]$ with linewidth given by $\gamma$. Here we assume that the response of the mechanical resonator is always linear. In practice, when the resonator is driven in large amplitude, mechanical nonlinearity such as Duffing effect could come into play. A discussion about the effect of mechanical nonlinearity in self-sustained optomechanical oscillation can be found in Ref. \cite{Tang_PRA_2012}. In this paper we will neglect the mechanical nonlinearity.

In summary, in this section we derive the equations for the coherent amplitude and small signal for the five output quantities: $\hat{a}$ (Eq. \eqref{eq:a_tfunc}), $\hat{s}_{out}$ (Eq. \eqref{eq:sout_tfunc}), $\hat{n}$ (Eq. \eqref{eq:n_tfunc}), $\hat{P}_{out}$ (Eq. \eqref{eq:Pout_tfunc}), and $\hat{x}$ (Eq. \eqref{eq:x_tfunc}). They form a set of equations which can be combined to solve for a self-consistent solution. 

\section{Results and discussions}\label{sec:results}

\subsection{Limit-cycle solution}\label{subsec:limit_cycle}

\begin{figure}[t]
\centering
\includegraphics[width=8.6cm]{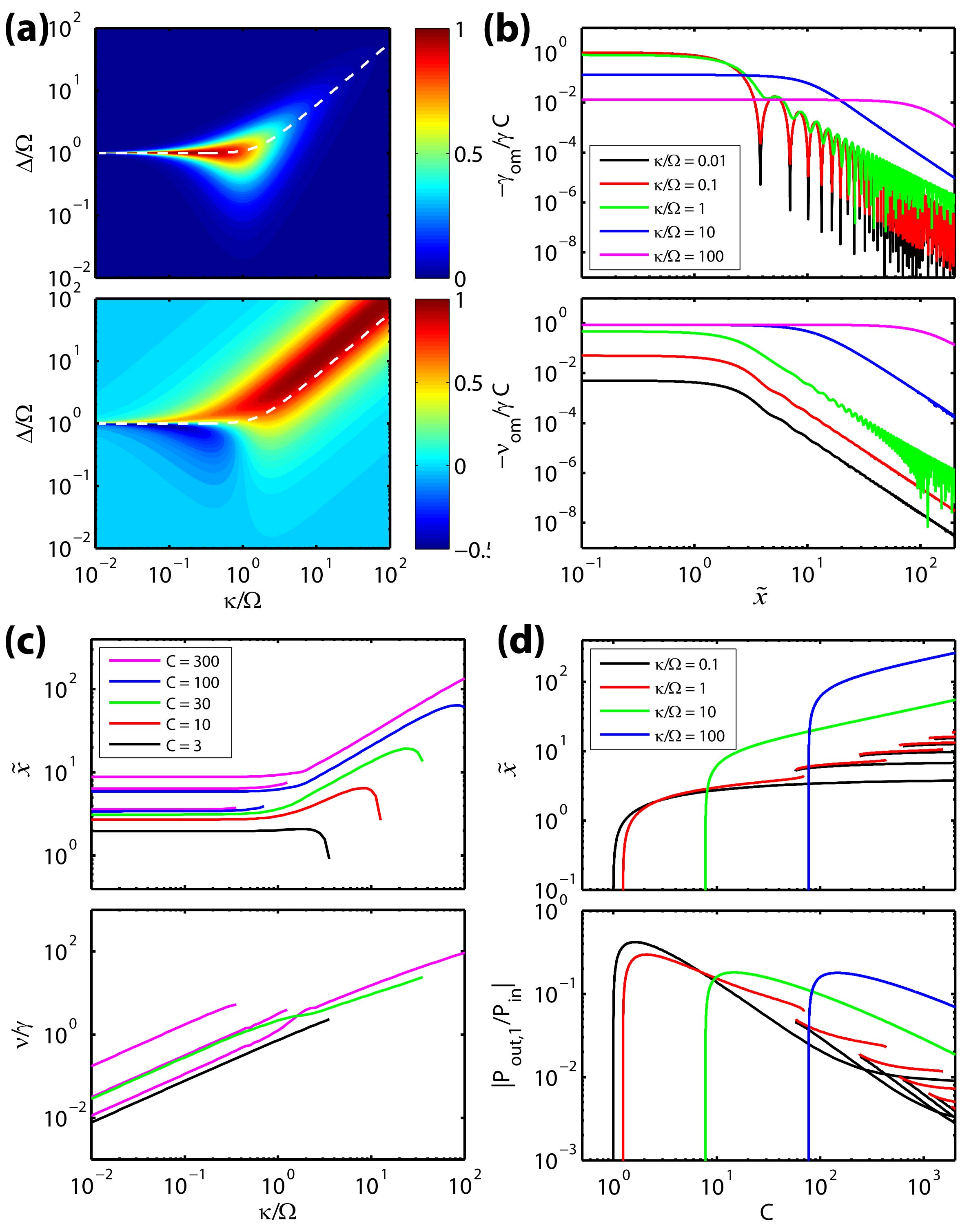}
\caption{ (a) Contour plot of (upper) $-\gamma_{om}(\tilde{x}=0)/\gamma C$ and (lower) $-\nu_{om}(\tilde{x}=0)/\gamma C$ as functions of $\Delta/\Omega$ and $\kappa/\Omega$. The white dashed lines indicate the optimal detuning where the optomechanical anti-damping at zero displacement $-\gamma_{om}(\tilde{x}=0)$ is maximum. (b) (upper) $-\gamma_{om}/\gamma C$ and (lower) $-\nu_{om}/\gamma C$ plotted as functions of $\tilde{x}$ at different $\kappa/\Omega$. (c) (upper) The limit-cycle amplitude $\tilde{x}$ and (lower) frequency shift $\nu/\gamma$ plotted as functions of $\kappa/\Omega$. (d) (upper) The limit-cycle amplitude $\tilde{x}$ and (lower) the normalized output power $|P_{out,1}/P_{in}|$ plotted as functions of $C$. In the calculations involved in (b), (c) and (d), optimal detuning described by the white dashed line of (a) is used.  }
\label{fig:LCamp}
\end{figure}

In this section, we put together the results derived in the previous section to solve for the closed-loop response of the system at the limit-cycle. 
A closed-form equation for the coherent amplitude can be obtained by combining the equations for $n_1$ (Eq. \eqref{eq:n_n}) and $x_1$ (Eq. \eqref{eq:x_1}). After $x_1$ is found, all the other harmonic components of $a_n$, $s_{out,n}$, $n_n$, and $P_{out,n}$ can be calculated using Eqs. \eqref{eq:a_n}, \eqref{eq:sout_n}, \eqref{eq:n_n}, \eqref{eq:Pout_n}. As mentioned above, here we only focus on three quantities: $x_1$, $n_1$, $P_{out,1}$. The equations are
\begin{align}
\label{eq:LC_n1}
n_1 &= \frac{\gamma C}{g}H(\tilde{x})x_1 \\
\label{eq:LC_x1}
x_1 &= \frac{ig}{\Gamma}n_1 \\
\label{eq:LC_P1}
P_{out,1} &= (i\Omega-2\kappa_i)n_1 ~,
\end{align}
where $C= n_0|_{x_1=0} (g^2/\kappa\gamma)$ is the cooperativity and $n_0|_{x_1=0}=2\kappa_e P_{in}/(\kappa^2+\Delta^2)$ is the static cavity photon number when $\tilde{x}=0$. We emphasize that $C$ is the cooperativity at zero displacement amplitude $\tilde{x}=0$. When the displacement amplitude is increased, the static photon number $n_0$ changes according to Eq. \eqref{eq:n_n} and so does the cooperativity. Nevertheless, $C$ is an important parameter determining when the self-oscillation starts. It is also an important parameter in optomechanical systems 
directly reflecting the strength of the laser drive. in Eq. \eqref{eq:LC_n1} the dimensionless function $H(\tilde{x})$ is defined as
\begin{equation}\label{eq:funcH}
H(\tilde{x})\equiv\frac{2\kappa^3}{\Omega} \left[1+\frac{\Delta^2}{\kappa^2}\right] \sum_m\frac{J_m(\tilde{x})J_{m+1}(\tilde{x})} {\tilde{x}K_m^\ast K_{m+1}} ~.
\end{equation}
 In the view of a feedback loop (see Figs. \ref{fig:feedback_loop} (a) and (b)), Eq. \eqref{eq:LC_n1} describes the response of the amplifier component and Eq. \eqref{eq:LC_x1} describes the response of the resonator. While the response of the resonator is assumed to be always linear, the amplifier has a amplitude-dependent gain described by the function $H(\tilde{x})$. Self-oscillation starts when the loop gain compensates the total loss. Note that the coherent amplitudes $n_1$ and $x_1$ are complex number, and so are $H(\tilde{x})$ and $\Gamma$. Therefore, Eqs. \eqref{eq:LC_n1} and \eqref{eq:LC_x1} describe an amplitude change as well as a phase shift caused by the corresponding component. A closed-loop solution is obtained by matching both the amplitude and phase. If we denote the real part and imaginary part of $H(\tilde{x})$ as
\begin{equation}\label{eq:gamma_nu_om}
\gamma C H(\tilde{x}) = \nu_{om}(\tilde{x}) +i\gamma_{om}(\tilde{x}) ~,
\end{equation}
the two equations \eqref{eq:LC_n1} and \eqref{eq:LC_x1} can be combined and rewritten as
\begin{equation}
[(\gamma+\gamma_{om}(\tilde{x})) -i(\nu+\nu_{om}(\tilde{x}))]x_1=0 ~.
\end{equation}
In order to have solution with non-zero $x_1$, we must have $\gamma=-\gamma_{om}$ and $\nu=-\nu_{om}$. This is essentially the Barkhausen condition for self-sustained oscillation, which states that at stable oscillation the loop gain is one and the loop phase shift is integral multiple of $2\pi$ \cite{Rubiola_Book_2010}. Therefore, we can interpret $-\nu_{om}$ as the frequency shift due to optical spring effect and $-\gamma_{om}$ as the optomechanical anti-damping rate. At the small-amplitude limit, i.e., $\tilde{x}\rightarrow0$, using the approximation $J_n(\tilde{x})\approx(\tilde{x}/2)^2/n!$, it can be shown that the results reduce back to the standard form in traditional analysis of optomechanical system. Fig. \ref{fig:LCamp} (a) shows the contour plots of $-\gamma_{om}(\tilde{x}=0)/\gamma C$ and $-\nu_{om}(\tilde{x}=0)/\gamma C$ as functions of $\kappa/\Omega$ and $\Delta/\Omega$. Only situations with positive $\Delta$ (blue-detuned) are shown. White-dashed line shows the optimal detuning at each given $\kappa/\Omega$ which gives the largest $-\gamma_{om}(\tilde{x}=0)/\gamma C$. Self-oscillation starts as long as $\gamma+\gamma_{om}\le 0$, i.e., when optomechanical anti-damping rate compensates the mechanical damping rate.

After the self-oscillation starts, the oscillation amplitude will grow until the nonlinearity kicks in. 
Fig. \ref{fig:LCamp} (b) shows $-\gamma_{om}/\gamma C$ and $-\nu_{om}/\gamma C$ versus $\tilde{x}$ at different $\kappa/\Omega$. In the calculation, optimal detuning (white dashed-line in Fig. \ref{fig:LCamp} (a)) at any given $\kappa$ is used. For the optomechanical anti-damping rate $-\gamma_{om}/\gamma C$, the curve is generally flat at small amplitudes and decays at large amplitude limit. In the RSR ($\kappa/\Omega<1$), the function is approximately one at small amplitude limit $\tilde{x}\rightarrow0$, implying that the oscillation starts as soon as $C>1$. The function rolls off at $\tilde{x}\gtrsim1$ and shows undulations with dips that go deeper for small $\kappa/\Omega$. This undulation in the gain function leads to multistability of the system \cite{Girvin_PRL_2006, Tang_PRA_2012}. In the USR ($\kappa/\Omega>1$), the flat regime has a smaller value, meaning that higher cooperativity $C$ is required to start self-oscillation. Also the regime extends until $\tilde{x}>\kappa/\Omega$. The optical spring effect $\mathrm{Re}\{H(\tilde{x})\}$ plotted in the lower panel of Fig. \ref{fig:LCamp} (b) shows similar features but with smaller undulations in the RSR.

With the knowledge of $H(\tilde{x})$, the steady-state amplitude $\tilde{x}$ and the mechanical detuning $\nu$ at the limit-cycle can be determined by invoking the Barkhausen condition. Fig. \ref{fig:LCamp} (c) plots $\tilde{x}$ and $\nu$ against $\kappa/\Omega$ at different $C$. Some of the curves end at large $\kappa/\Omega$ when $C$ is lower than the threshold value and so the system is not self-oscillating. In the RSR, the multistability manifests itself as multiple stable amplitudes, with more branches appearing for larger $C$ or smaller $\kappa/\Omega$. One can observe a general trend that in the RSR the values of $\tilde{x}$ stay more or less the same due to the similarity in $\gamma_{om}/\gamma C$. In the USR, a larger $C$ is required to start oscillations for larger $\kappa/\Omega$, but once it starts the amplitude leaps to larger value due to the wider flat regime in $\gamma_{om}/\gamma C$. For the mechanical detuning $\nu$ a general trend is that it increases with $\kappa/\Omega$ and becomes larger than the mechanical linewidth $\gamma$ in the USR.

The upper panel of Fig. \ref{fig:LCamp} (d) plots $\tilde{x}$ as a function of $C$. It corresponds to the situation when the input laser power is increased. Once $C$ is larger than a threshold value, the steady state amplitude rises sharply and then saturates. This threshold behavior of the optomechanical oscillation has been compared analogously with the lasing phenomenon \cite{Rabinovich_PRL_2012}. Note that $\tilde{x}$ is the displacement amplitude but not the oscillator signal that is detected. Using the Eqs.~\eqref{eq:Pout_n}, the detected output power is given by
\begin{equation}
\left|\frac{P_{out,1}}{P_{in}}\right| =\frac{\kappa_e\Omega\sqrt{\Omega^2+4\kappa_i^2}} {\kappa(\Delta^2+\kappa^2)} |H(\tilde{x})|\tilde{x} ~.
\end{equation}
The lower panel of Fig. \ref{fig:LCamp} (d) plots $|P_{out,1}/P_{in}|$ as a function of $C$ with critical coupling condition $\kappa_e=\kappa_i$ assumed.

\subsection{Closed-loop noise response}\label{subsec:noise_response}

After solving for the coherent amplitudes, they can be substitution into Eqs. \eqref{eq:da_n}, \eqref{eq:dsout_n}, \eqref{eq:dn_n}, \eqref{eq:dPout_n}, and \eqref{eq:dx_1} to solve for the closed-loop noise response. A closed-form equation can be obtained by combining the equations for $\delta\hat{n}_1$ and $\delta\hat{x}_1$,
\begin{equation}\label{eq:LC_dn_1}
\begin{aligned}
&\delta\hat{n}_1[\omega] = \frac{\sqrt{2\gamma}}{g}\delta\hat{f}_{op,1} \\
&~~+2\kappa_e P_{in} \sum_m
\frac{ig J_m^2}{|K_m|^2} \left[\frac{L_{K_{m+1}}}{K_{m+1}}-\frac{L_{K_{m-1}^\ast}}{K_{m-1}^\ast}\right] \delta\hat{x}_1 \\
&~~+2\kappa_e P_{in} \sum_m
\frac{ig J_{m-1}J_{m+1}}{K_m^\ast K_{m+1}} \left[\frac{L_{K_{m}}}{K_{m}}-\frac{L_{K_{m}^\ast}}{K_{m}^\ast}\right] \delta\hat{x}_1^\ddagger \\
&\delta\hat{x}_1[\omega] = \frac{iL_{\Gamma}[\omega]}{\Gamma} (g\delta\hat{n}_1[\omega]+\sqrt{2\gamma}\delta\hat{f}_{th,1}[\omega]) ~.
\end{aligned}
\end{equation}
Here we have grouped the terms due to laser noise and cavity vacuum fluctuation as
\begin{equation}\label{eq:df_op}
\begin{aligned}
&\delta\hat{f}_{op,1}=\sqrt{\gamma C 2\kappa_e\kappa}\\
&\sum_{m,p}\left\{ \frac{|K_0|J_{m-1}J_{m-p}L_{K_m}}{K_{m-1}^\ast K_m} \left[\delta\hat{s}_{in,p} +\sqrt{\frac{\kappa_i}{\kappa_e}}\delta\hat{s}_{vac,p}\right]\right. \\
&\left. + \frac{|K_0|J_{m+1}J_{m-p}L_{K_m^\ast}}{K_{m+1} K_m^\ast} \left[\delta\hat{s}_{in,p}^\ddagger +\sqrt{\frac{\kappa_i}{\kappa_e}}\delta\hat{s}_{vac,p}^\ddagger\right]\right\} ~.
\end{aligned}
\end{equation}
This term is normalized to have the same unit as the thermal fluctuation force $\delta\hat{f}_{th,1}$. When the laser noise is quantum noise limited, this term represents the optical force due to vacuum fluctuations and is the radiation pressure shot noise \cite{Regal_Science_2013}.
\begin{figure}[t]
\centering
\includegraphics[width=7cm]{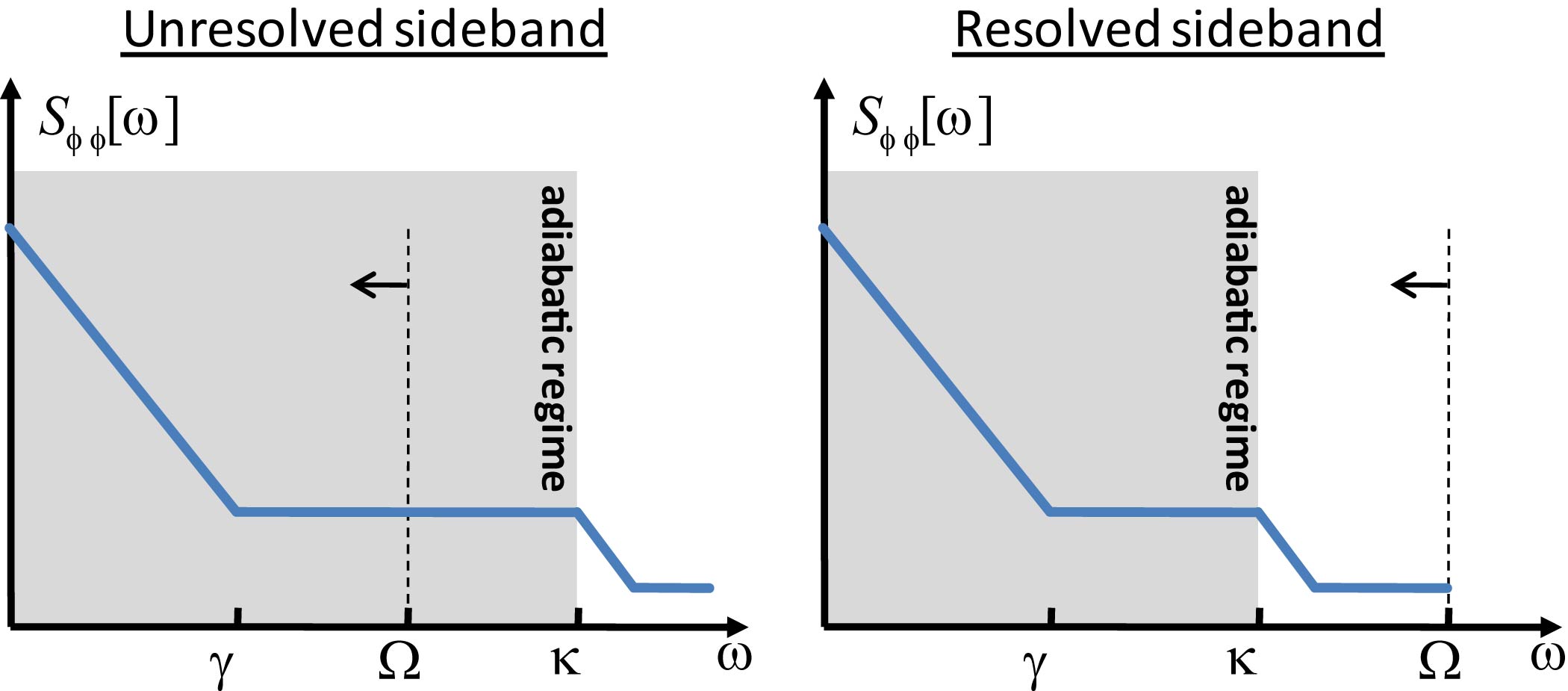}
\caption{ Schematic illustrating the adiabaticity condition. The area shaded with the grey color indicates the frequency range $\omega<\kappa$ where adiabatic assumption is valid. The dashed-line and the arrow indicate the frequency range $\omega<\Omega/2$ where $\omega$ is defined. In the unresolved sideband regime the adiabatic assumption is always valid. }
\label{fig:adiab_regime}
\end{figure}

\begin{figure}
\centering
\includegraphics[width=8.6cm]{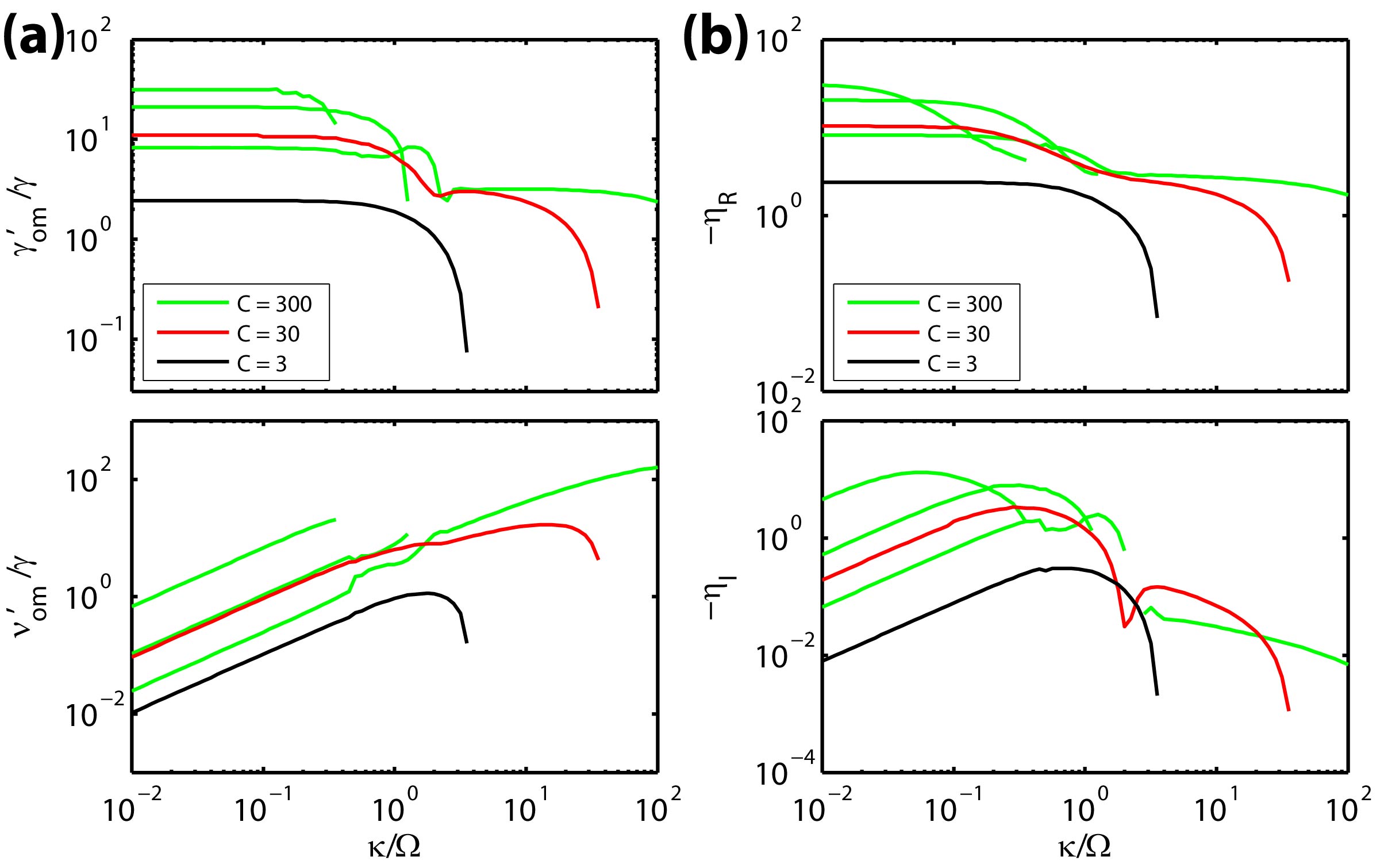}
\caption{ (a) (upper) $\gamma_{om}^\prime/\gamma$ and (lower) $\nu_{om}^\prime/\gamma$ plotted as a function of $\kappa/\Omega$. (b) (upper) $-\eta_R$ and (lower) $-\eta_I$ plotted as a function of $\kappa/\Omega$. }
\label{fig:LCnoise}
\end{figure}

In principle, Eq. \eqref{eq:LC_dn_1} can be solved by brute-force numerical calculations. Instead of taking this approach, we first simplify the equation by considering only the frequency range $\omega<\kappa$. For the USR, this condition is always satisfied since from the definition of the spectral filter decomposition $\omega$ is always smaller than $\Omega/2$, which in turn is smaller than $\kappa$. Even for the RSR, the condition is also valid for a range of frequencies which includes $\omega<\gamma$ since in most practical cases the mechanical linewidth is much smaller than optical cavity linewidth ($\gamma\ll\kappa$). $\omega<\gamma$ is the frequency range of important interest because it is where the mechanical resonator has a significant response. The assumption $\omega<\kappa$ means that the time scale under consideration is longer than the optical response time. 
We call this condition ``adiabatic regime'' since effectively the optical cavity responds instantly. Fig. \ref{fig:adiab_regime} shows a schematic illustrating the adiabatic condition in the USR and the RSR. Mathematically, the condition $\omega<\kappa$ means we can take all the optical filter functions $L_{K_m}[\omega\rightarrow0]$ to be $1$. With this approximation, it can be shown that the coefficients of the terms $\delta\hat{x}_1$ and $\delta\hat{x}_1^\ddagger$ can be written in terms of $H(\tilde{x})$ and its derivative $H^\prime(\tilde{x})$ as
\begin{equation}\label{Eq:adiab_dn1}
\begin{aligned}
&\delta\hat{n}_1 = \frac{\sqrt{2\gamma}}{g}\delta\hat{f}_{op,1}
+\frac{\gamma C}{g} \left[H(\tilde{x}) +\frac{H^\prime(\tilde{x})\tilde{x}}{2}\right] \delta\hat{x}_1 \\
&+\frac{\gamma C}{g} \left[\frac{H^\prime(\tilde{x})\tilde{x}}{2}\right] \delta\hat{x}_1^\ddagger\\
&\delta\hat{x}_1[\omega] = \frac{iL_{\Gamma}[\omega]}{\Gamma} (g\delta\hat{n}_1[\omega]+\sqrt{2\gamma}\delta\hat{f}_{th,1}[\omega]) ~.
\end{aligned}
\end{equation}
Similar to Eq. \eqref{eq:gamma_nu_om}, we denote the real part and imaginary part of $H^\prime(\tilde{x})\tilde{x}$ as
\begin{equation}\label{eq:Gamma_om_prime}
\gamma C H^\prime(\tilde{x})\tilde{x} = \nu_{om}^\prime(\tilde{x}) +i\gamma_{om}^\prime(\tilde{x}) ~.
\end{equation}
This quantity represents the effect of amplitude-dependence of the gain function. With this simplified form, it is can be shown that
\begin{equation}\label{eq:adiab_dx_A_phi}
\left[\begin{array}{c}
\delta\hat{x}_1^A \\ \delta\hat{x}_1^\phi
\end{array}\right]
=\frac{\sqrt{2\gamma}}{x_1}
\left[\begin{array}{cc}
\frac{1}{\gamma_{om}^\prime-i\omega} & 0 \\
\frac{i\nu_{om}^\prime}{\omega(\gamma_{om}^\prime-i\omega)} & \frac{i}{\omega}
\end{array}\right]
\left[\begin{array}{c}
\delta\hat{f}_{th}^A +\delta\hat{f}_{op}^A \\
\delta\hat{f}_{th}^\phi +\delta\hat{f}_{op}^\phi
\end{array}\right] ~,
\end{equation}
where $\delta\hat{f}^A=i(\delta\hat{f}_1 -\delta\hat{f}_1^\ddagger)/2$ and $\delta\hat{f}^\phi=(\delta\hat{f}_1 +\delta\hat{f}_1^\ddagger)/2$ are the forces acting in the amplitude and phase direction with respect to the displacement $x_1$. (A $\pi/2$ phase rotation is included to account for the phase lag between the force and the displacement when the resonator is driven at resonance. The subscript 1 is dropped for brevity.) Eq. \eqref{eq:adiab_dx_A_phi} is a closed-form expression relating the displacement noise to the thermal and optical force noise through a transfer matrix. From the expression, we can see that $\gamma_{om}^\prime$ is the linewidth of the amplitude noise and thus the damping rate of the amplitude fluctuations \cite{Tang_PRA_2012, Armour_PRL_2010}. $\nu_{om}^\prime$ represents the frequency shift due to amplitude change at the limit-cycle. 
Fig. \ref{fig:LCnoise} (a) plots $\gamma_{om}^\prime$ and $\nu_{om}^\prime$ in various situations.

From the equation, we can see that the phase fluctuation $\delta\hat{x}_1^\phi$ has an overall $1/\omega$ dependence, which leads to a $1/\omega^2$ dependence in the phase noise spectral density. Such a signature of phase diffusion is expected since for self-sustained oscillations there is no restoring constraint for the phase fluctuations. It is in contrary to the situation for amplitude fluctuations where the amplitude dependent gain function imposes a limit on the steady state amplitude. Mathematically, this overall $1/\omega$ dependence shows up as long as the transfer matrices relating the amplitude and phase noise of $\delta\hat{n}_1$ and $\delta\hat{x}_1$
\begin{align}
\label{eq:tmat_dx2dn}
\left[\begin{array}{c}
\delta\hat{n}_1^A \\ \delta\hat{n}_1^\phi
\end{array}\right]
&=
\left[\begin{array}{cc}
1+\mathrm{Re}\{\frac{H'\tilde{x}}{H}\} & 0 \\
\mathrm{Im}\{\frac{H'\tilde{x}}{H}\} & 1
\end{array}\right]
\left[\begin{array}{c}
\delta\hat{x}_1^A \\ \delta\hat{x}_1^\phi
\end{array}\right] +\ldots
\\
\label{eq:tmat_dn2dx}
\left[\begin{array}{c}
\delta\hat{x}_1^A \\ \delta\hat{x}_1^\phi
\end{array}\right]
&=
\left[\begin{array}{cc}
\frac{L_{\Gamma}+L_{\Gamma^\ast}}{2} & -\frac{L_{\Gamma}-L_{\Gamma^\ast}}{2i} \\
\frac{L_{\Gamma}-L_{\Gamma^\ast}}{2i} & \frac{L_{\Gamma}+L_{\Gamma^\ast}}{2}
\end{array}\right]
\left[\begin{array}{c}
\delta\hat{n}_1^A \\ \delta\hat{n}_1^\phi
\end{array}\right] +\ldots ~,
\end{align}
which can be obtained from Eqs. \eqref{Eq:adiab_dn1}, have the upper-right and lower-right matrix elements equal to 0 and 1 at the static limit $\omega\rightarrow0$. ($L_\Gamma[\omega\rightarrow0]=1$.) The physical meaning of this is that, at the static limit when all the transient response of the system has died out, a shift in phase of the input quantity will not affect the amplitude of the output quantity (0 in the upper-right matrix element) and will shift the phase of the output quantity by exactly the same amount (1 in the lower-right matrix element). These two conditions are satisfied as long as the system is time-translational invariant, since a shift in phase is equivalent to a shift in time which leaves the system unchanged. Therefore, any closed-loop time-translational invariant system always has a $1/\omega^2$ dependence in phase noise spectral density at the static limit.

Eq. \eqref{eq:adiab_dx_A_phi} can be substituted back to Eq. \eqref{Eq:adiab_dn1} to obtain a closed-form expression for $\delta\hat{n}_1$, which is given by
\begin{equation}\label{eq:adiab_dn_A_phi}
\begin{aligned}
\left[\begin{array}{c}
\delta\hat{n}_1^A \\ \delta\hat{n}_1^\phi
\end{array}\right]
&=\frac{\sqrt{2\gamma}}{x_1}
\left[\begin{array}{cc}
\frac{1+\eta_R}{\gamma_{om}^\prime-i\omega} & 0 \\
\frac{\eta_I \omega+i\nu_{om}^\prime}{\omega(\gamma_{om}^\prime-i\omega)} & \frac{i}{\omega}
\end{array}\right]
\left[\begin{array}{c}
\delta\hat{f}_{th}^A +\delta\hat{f}_{op}^A\\
\delta\hat{f}_{th}^\phi +\delta\hat{f}_{op}^\phi
\end{array}\right] \\
&+\frac{\sqrt{2\gamma}}{x_1(\gamma^2+\nu^2)}
\left[\begin{array}{cc}
\gamma & -\nu \\
\nu & \gamma
\end{array}\right]
\left[\begin{array}{c}
\delta\hat{f}_{op}^A \\
\delta\hat{f}_{op}^\phi
\end{array}\right] ~,
\end{aligned}
\end{equation}
where $\eta_R =\mathrm{Re} \{H^\prime\tilde{x}/H\}$ and $\eta_I =\mathrm{Im} \{H^\prime\tilde{x}/H\}$. From Eq. \eqref{eq:tmat_dx2dn} we can see that $\eta_R$ and $\eta_I$ account for the transfers from displacement amplitude noise $\delta\hat{x}_1^A$ to the photon number amplitude noise $\delta\hat{n}_1^A$ and phase noise $\delta\hat{n}_1^\phi$. A plot of $\eta_R$ and $\eta_I$ is shown in Fig. \ref{fig:LCnoise} (b). In Eq. \eqref{eq:adiab_dn_A_phi}, $\delta\hat{f}_{op}$ comes in two separate terms, one originates directly from the input laser and vacuum noise (first term in Eq. \eqref{Eq:adiab_dn1}) and the other is due to the feedback from the mechanical displacement.

Applying Eq. \eqref{eq:adiab_dn_A_phi} to Eq. \eqref{eq:dPout_n}, it can be shown that in adiabatic limit $\omega<\kappa$ the amplitude and phase noise of $\delta\hat{P}_{out,1}$ are given by
\begin{equation}\label{eq:adiab_dPout_A_phi}
\begin{aligned}
&\left[\begin{array}{c}
\delta\hat{P}_{out,1}^A \\ \delta\hat{P}_{out,1}^\phi
\end{array}\right]
=
\left[\begin{array}{cc}
1 & 0 \\
0 & 1
\end{array}\right]
\left[\begin{array}{c}
\delta\hat{n}_1^A\\
\delta\hat{n}_1^\phi
\end{array}\right]
+
\left[\begin{array}{c}
\delta\hat{P}_{vac}^A \\
\delta\hat{P}_{vac}^\phi
\end{array}\right]
\\
&~~~~+
\left[\begin{array}{cc}
\mathrm{Re}(\frac{P_{in}}{P_{out,1}}) & -\mathrm{Im}(\frac{P_{in}}{P_{out,1}}) \\
\mathrm{Im}(\frac{P_{in}}{P_{out,1}}) & \mathrm{Re}(\frac{P_{in}}{P_{out,1}})
\end{array}\right]
\left[\begin{array}{c}
\delta\hat{P}_{in,1}^A \\
\delta\hat{P}_{in,1}^\phi
\end{array}\right] ~,
\end{aligned}
\end{equation}
where the second term $\delta\hat{P}_{vac}$ take into account vacuum noise and the last term is the input laser noise that goes directly to the output.

Eqs. \eqref{eq:adiab_dx_A_phi}, \eqref{eq:adiab_dn_A_phi}, and \eqref{eq:adiab_dPout_A_phi} are the closed-form expressions that can be used to directly calculate the amplitude and phase noise for $\delta\hat{x}_1$, $\delta\hat{n}_1$, and $\delta\hat{P}_{out,1}$ (see the highlighted region in Fig. \ref{fig:spectrum}). These expressions are valid only for adiabatic limit, which is always valid in the USR. In the RSR when $\omega$ significantly deviates from the adiabatic condition, one can use the full expressions in Eq. \eqref{eq:LC_dn_1} for a full calculation. In the following sections, we will examine individually the phase noise from three contributions: thermomechanical noise, photon shot noise, and low-frequency technical laser noise. Since these three noise sources are uncorrelated, the combined phase noise will be the sum of the individual contribution.

\subsection{Phase noise contribution from thermomechanical fluctuation}\label{subsec:phasenoise_thermal}

We first look at the phase noise due to thermomechanical noise. In this case we neglect all the terms with $\delta\hat{s}_{in}$ and $\delta\hat{s}_{vac}$ and consider only the effect of $\delta\hat{f}_{th}$. We define the correlation matrix in frequency domain as
\begin{equation}\label{eq:corr_matrix_def}
\mathcal{C}_{\delta\hat{q}}[\omega]=
\left[\begin{array}{cc}
\mathcal{S}_{\delta\hat{q}^A \delta\hat{q}^A} [\omega] & \mathcal{S}_{\delta\hat{q}^A \delta\hat{q}^\phi} [\omega] \\
\mathcal{S}_{\delta\hat{q}^\phi \delta\hat{q}^A} [\omega] & \mathcal{S}_{\delta\hat{q}^\phi \delta\hat{q}^\phi} [\omega]
\end{array}\right] ~,
\end{equation}
%
It can be shown that the thermomechanical force noise correlation matrix is given by
\begin{equation}\label{eq:corr_matrix_dfth}
\mathcal{C}_{\delta\hat{f}_{th}}[\omega]= \frac{1}{2}
\left[\begin{array}{cc}
\bar{n}_{th} + \frac{1}{2} & \frac{i}{2} \\
\frac{-i}{2} & \bar{n}_{th} + \frac{1}{2}
\end{array}\right] ~.
\end{equation}
The single-sided noise spectral density accessible in experiments is defined as $\mathcal{S}_{\hat{X}}^{SS}[\omega] =\mathcal{S}_{\hat{X}\hat{X}}[\omega] +\mathcal{S}_{\hat{X}\hat{X}}[-\omega]$. In the adiabatic limit, the expressions for the amplitude and phase noise of $\delta\hat{x}_1$ can be obtained directly from Eqs.~\eqref{eq:adiab_dx_A_phi} and \eqref{eq:corr_matrix_dfth}, and are given by
\begin{align}
\label{eq:Sxx_amp_th}
\mathcal{S}_{\delta\hat{x}_1^A}^{SS} [\omega] &= \frac{\gamma^2}{\gamma_{om}^{\prime 2}+\omega^2} \frac{2\bar{n}_{th}+1}{\gamma n_{x}} \\
\label{eq:Sxx__ph_th}
\mathcal{S}_{\delta\hat{x}_1^\phi}^{SS} [\omega] &= \left[\frac{\gamma^2}{\omega^2} +\frac{\nu_{om}^{\prime2}}{\omega^2} \frac{\gamma^2}{\gamma_{om}^{\prime 2}+\omega^2} \right] \frac{2\bar{n}_{th}+1}{\gamma n_{x}} ~,
\end{align}
where $n_x=|x_1|^2$ is the phonon number calculated from the coherent amplitude. As expected, the relative amplitude noise and phase noise is proportional to the ratio between the thermal phonon number $\bar{n}_{th}$ and the coherent phonon number $n_x$. We can see that the relative amplitude noise spectrum is a Lorentzian function with linewidth of $\gamma_{om}^{\prime}$, while the phase noise spectrum consists of two terms. The first term is the typical $1/\omega^2$ term accounts for the phase diffusion. The second term can be attributed to the amplitude-noise-induced-frequency-noise through the optical spring effect: amplitude fluctuations modify the oscillation frequency through the optical spring effect ($\nu$ is a function of $\tilde{x}$), which in turn manifest as phase noise since $\omega_{osc}=d\phi/dt$. As a result this term is equal to $\nu_{om}^{\prime2}/\omega^2$ multiplied by the amplitude noise spectral density.

\begin{figure}
\centering
\includegraphics[width=8.6cm]{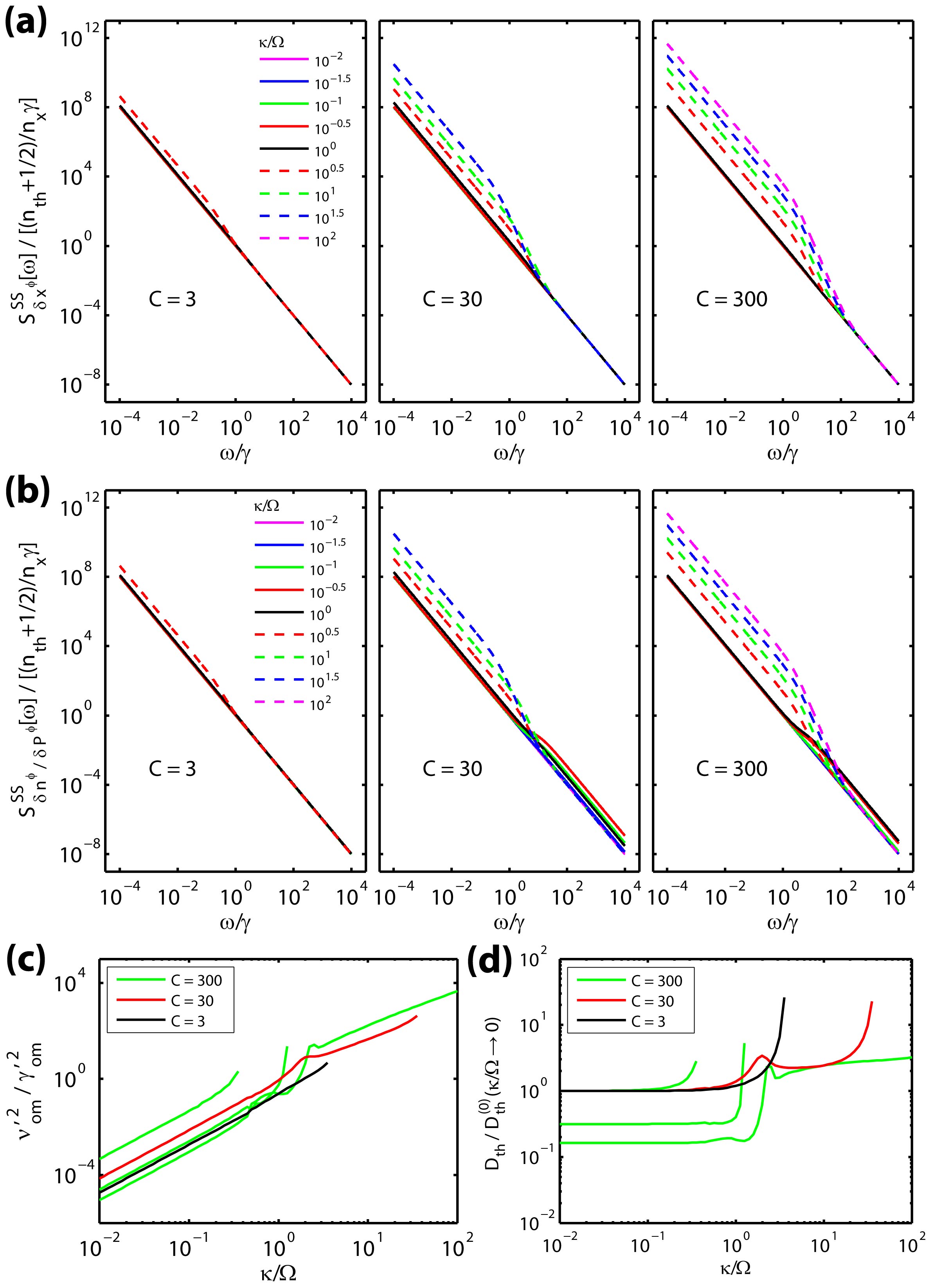}
\caption{ Phase noise spectral densities of (a) $\delta\hat{x}_1$ and (b) $\delta\hat{n}_1$ and $\delta\hat{P}_{out,1}$ due to thermomechanical noise.  Solid lines are used for the RSR and dashed lines are used for the USR. (c) $\nu_{om}^{\prime 2}/\gamma_{om}^{\prime 2}$ plotted against $\kappa/\Omega$. (d) Normalized phase diffusion constant $\mathcal{D}_{th} / \mathcal{D}_{th}^{(0)} (\kappa/\Omega\rightarrow 0)$ plotted against $\kappa/\Omega$. $\mathcal{D}_{th}^{(0)}$ denotes the diffusion constant for the limit-cycle with the smallest displacement when multiple solutions exist. }
\label{fig:phasenoise_thermal}
\end{figure}

For the phase noise spectral densities of $\delta\hat{n}_1^\phi$ and $\delta\hat{P}_{out,1}^\phi$, it can be shown that they have the same expression and are given by
\begin{equation}
\begin{aligned}
\label{eq:Snn_SPP_ph_th}
&\mathcal{S}_{\delta\hat{n}_1^\phi}^{SS} [\omega] =\mathcal{S}_{\delta\hat{P}_{out,1}^\phi}^{SS} [\omega] \\
&= \left[\frac{\gamma^2}{\omega^2} +\frac{\nu_{om}^{\prime2}}{\omega^2} \frac{\gamma^2}{\gamma_{om}^{\prime 2}+\omega^2} +\frac{\eta_I^2\gamma^2}{\gamma_{om}^{\prime 2}+\omega^2} \right] \frac{2\bar{n}_{th}+1}{\gamma n_{x}} .
\end{aligned}
\end{equation}
Compared to Eq. \eqref{eq:Sxx__ph_th}, there is an additional term proportional to $\eta_I^2$ which is due to the amplitude-noise-to-phase-noise transfer (see Eq. \eqref{eq:tmat_dx2dn}).

Fig. \ref{fig:phasenoise_thermal} (a) and (b) plot the phase noise spectral densities calculated from Eqs. \eqref{eq:Sxx__ph_th} and \eqref{eq:Snn_SPP_ph_th} normalized with $(\bar{n}_{th}+1/2)/n_x\gamma$ for different $\kappa/\Omega$ and $C$. The spectra are calculated based on the limit-cycle amplitude obtained in the previous sections. For the RSR where there are multiple branches of solutions, the solution with smallest displacement amplitude is used. In the plots with $C=3$ and $C=30$, some of the curves with large $\kappa/\Omega$ are absent because the cooperativity is below threshold and hence the device is not self-oscillating. From the figure, we can see that in general, unlike the commonly known ``L-shape'' in oscillator phase noise spectra \cite{Rubiola_Book_2010}, the phase noise due to thermomechanical noise are decreasing functions through the whole frequency range. This is expected since the noise enters the system through the mechanical resonator, which at the first place filters out the noise that is beyond the resonator linewidth. We can also see that there is a significant rise of phase noise for the USR at the low frequency side, which is due to the larger $\nu_{om}^\prime$ in this regime (see Fig. \ref{fig:LCnoise} (a)). On the other hand, in the RSR the larger $\eta_I$ (see Fig. \ref{fig:LCnoise} (b)) adds more phase noise to $\delta\hat{n}_1$ and $\delta\hat{P}_{out,1}$ in the high frequency side.

The $1/\omega^2$ dependence in noise spectrum is a signature of a diffusion process (or random walk). The phase diffusion constant $\mathcal{D}$ can be obtained from the coefficient of the $1/\omega^2$ term at low frequency limit $\omega\rightarrow0$. For phase noise due to thermomechanical fluctuation, it is given by
\begin{equation}\label{eq:Dth}
\mathcal{D}_{th} = \gamma\frac{\bar{n}_{th}+\frac{1}{2}}{n_x} \left[1+\frac{\nu_{om}^{\prime2}}{\gamma_{om}^{\prime2}}\right] ~.
\end{equation}
This expression is consistent with the result in Ref. \cite{Armour_PRL_2010}. The behavior of the factor $\nu_{om}^{\prime2}/\gamma_{om}^{\prime2}$ is plotted in Fig. \ref{fig:phasenoise_thermal} (c), which shows that it has a significantly higher value in the USR than in the RSR. Yet, in the USR the displacement amplitude is larger (see Fig. \ref{fig:LCamp} (e)) and so the thermal phonon to coherent phonon ratio is smaller. With both effects taken into account, Fig. \ref{fig:phasenoise_thermal} plots $\mathcal{D}_{th}$ against $\kappa/\Omega$ at different $C$. At each value of $C$, the plotted value is normalized with the diffusion constant calculated at the limit of $\kappa/\Omega\rightarrow0$ (among the multi-solutions in the RSR, the solution with the smallest displacement is used). We can see that even when we take into account the displacement amplitude difference, the phase diffusion constant in the RSR is still generally smaller than that in the USR.

\subsection{Phase noise contribution from photon shot noise }\label{subsec:phasenoise_shot}

Next, we consider the phase noise contribution from the photon shot noise. 
In this case we take $\delta\hat{f}_{th}$ to be zero and set the correlators of $\delta\hat{s}_{in}$ to be the same as that of $\delta\hat{s}_{vac}$. Using the expression for the optical force noise $\delta\hat{f}_{op}$ in Eq. \eqref{eq:df_op} and the vacuum noise correlators in Eq. \eqref{eq:noise_corr_vac}, the optical force correlation matrix can be found as
\begin{equation}\label{eq:corr_matrix_dfsh}
\mathcal{C}_{\delta\hat{f}_{vac}}[\omega]=C
\left[\begin{array}{cc}
F_{-} & F_{R}+iF_{I} \\
F_{R}-iF_{I} & F_{+}
\end{array}\right] ~,
\end{equation}
where
\begin{align}\label{eq:FpmRI}
&F_{\pm}[\omega]=\sum_m \left|\frac{J_{m-1}}{K_{m-1}} \pm\frac{J_{m+1}}{K_{m+1}}\right|^2 \frac{\kappa^2|K_0|^2}{4|K_m|^2} \left|L_{K_m}[\omega]\right|^2 \\
&F_{R/I}[\omega]=\mathrm{Re/Im}\left\{ i\sum_m \left[\frac{J_{m-1}}{K_{m-1}} -\frac{J_{m+1}}{K_{m+1}}\right]^\ast\right. \nonumber \\
&~~~~~~\times \left. \left[\frac{J_{m-1}}{K_{m-1}} +\frac{J_{m+1}}{K_{m+1}}\right] \frac{\kappa^2|K_0|^2}{4|K_m|^2} \left|L_{K_m}[\omega]\right|^2\right\} ~.
\end{align}
%
As mentioned before, this optical force is due to vacuum fluctuations and is essentially the radiation pressure shot noise \cite{Regal_Science_2013}. This optical force noise is proportional to cooperativity $C$ and in the case of self-sustained oscillations there is an implicit $C$ dependence in the functions $F_{\pm}$ and $F_{R/I}$ since the limit-cycle amplitude depends on $C$. One interesting feature is that, unlike the thermomechanical force noise, this optical force noise is in general not isotropic, i.e., the forces acting on different quadrature directions can be different. The principle axes and the corresponding force noise correlation are given by the eigenvectors and eigenvalues of the correlation matrix. 
Fig. \ref{fig:phasenoise_shot}~(a) and (b) plots the $|F_\pm|$ and $|F_{R/I}|$ for the adiabatic limit $\kappa\ll\omega$. 

\begin{figure}
\centering
\includegraphics[width=8.6cm]{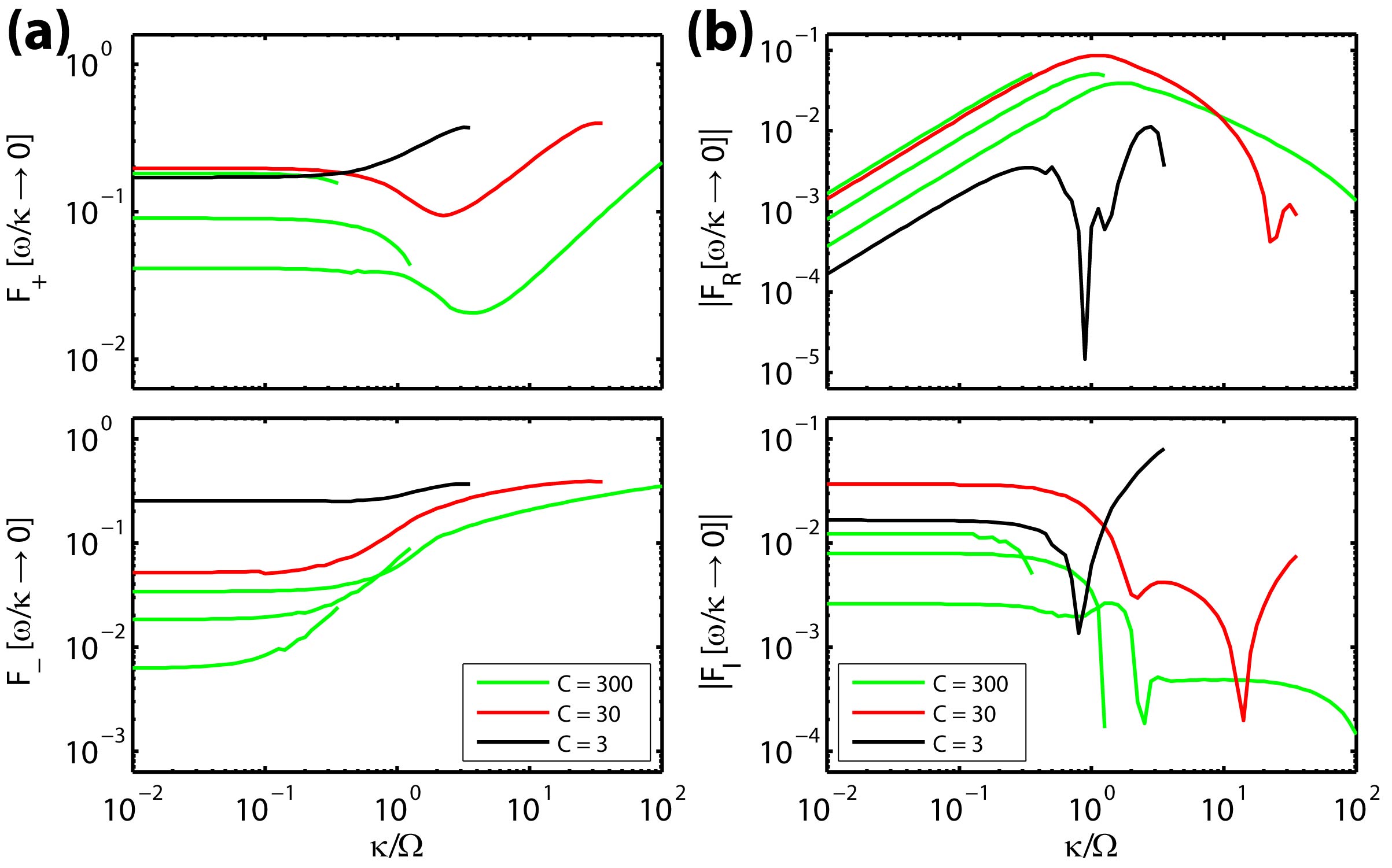}
\caption{ Correlation matrix elements of the optical shot noise force (Eq. \eqref{eq:corr_matrix_dfsh}) in the adiabatic limit plotted against $\kappa/\Omega$. }
\label{fig:dfvc_noise}
\end{figure}

With the optical force noise correlation matrix $\mathcal{C}_{\delta\hat{f}_{vac}}$, the phase noise spectral density of $\delta\hat{x}_1$ can be obtained using the transfer matrix in Eq. \eqref{eq:adiab_dx_A_phi}, and is given by
\begin{equation}\label{eq:Sxx_sh}
\begin{aligned}
\mathcal{S}_{\delta\hat{x}_1^\phi}^{SS} [\omega] &= \frac{\gamma^2}{\omega^2}\left[F_+ + \frac{\nu_{om}^{\prime2}F_- +2\gamma_{om}^\prime\nu_{om}^\prime F_R} {\gamma_{om}^{\prime 2}+\omega^2} \right] \frac{4C}{\gamma n_{x}} ~.
\end{aligned}
\end{equation}
Similar to Eq. \eqref{eq:Sxx__ph_th}, the phase noise spectrum consists of two terms, one from the phase diffusion and one from the amplitude-noise-induced-frequency-noise. For the latter term, there is an additional contribution from the amplitude-phase correlation $F_R$. Fig. \ref{fig:phasenoise_shot} (a) plots $\mathcal{S}_{\delta\hat{x}_1^\phi}^{SS} [\omega]$ at different $\kappa/\Omega$ and $C$. It shows similar features as in the case for phase noise due to thermomechanical noise.

\begin{figure}
\centering
\includegraphics[width=8.6cm]{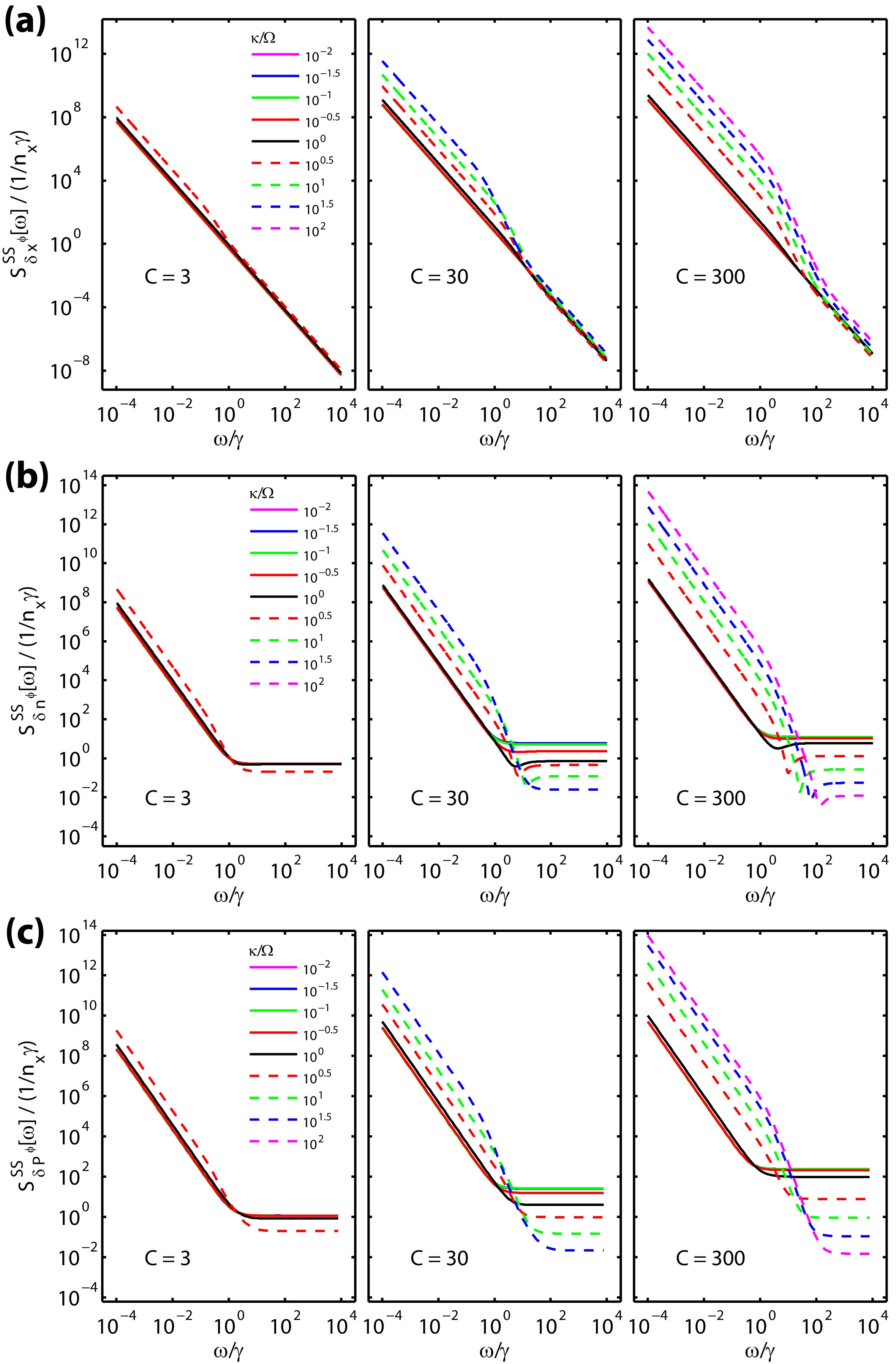}
\caption{ Phase noise spectral densities of (a) $\delta\hat{x}_1$ and (b) $\delta\hat{n}_1$ and (c) $\delta\hat{P}_{out,1}$ due to photon shot noise. $\gamma/\Omega=10^{-4}$ and $\kappa_e=\kappa_i$ is assumed in the calculation. Solid lines are used for the RSR and dashed lines are used for the USR. }
\label{fig:phasenoise_shot}
\end{figure}

The phase noise spectral density of $\delta\hat{n}_1$ can be obtained using the transfer matrix in Eq. \eqref{eq:adiab_dx_A_phi}. As we mentioned before, in Eq. \eqref{eq:adiab_dx_A_phi} $\delta\hat{f}_{op}$ comes in two separate terms, one originates directly from the input laser and vacuum noise (first term in Eq. \eqref{Eq:adiab_dn1}) and the other is from the mechanical displacement. While the latter is expected to be decaying as $\gamma^2/\omega^2$, the former gives a relatively flat background which only starts decreasing outside the adiabatic region, where $\omega>\kappa$. When these two terms add up, it gives to the commonly seen ``L-shape'' phase noise spectrum. 
The full expression for the phase noise spectral density of $\delta\hat{n}_1$ is given by
\begin{equation}\label{eq:Snn_sh}
\begin{aligned}
&\mathcal{S}_{\delta\hat{n}_1^\phi}^{SS} [\omega] = \\
&\left\{\left[\frac{\gamma^2}{\omega^2} + \frac{\gamma^4}{(\gamma^2+\nu^2)^2}\right] \left[F_+ + \frac{F_- \nu_{om}^{\prime2} +2F_R\nu_{om}^\prime\gamma_{om}^\prime} {\gamma_{om}^{\prime 2}+\omega^2} \right] \right.\\
&~~~~+\frac{\gamma^2\nu\omega^2 (\nu F_- + 2\gamma F_R)} {(\gamma^2+\nu^2)^2 (\gamma_{om}^{\prime 2}+\omega^2)} \\
&~~~~\left.- \frac{2\gamma^2\nu (\nu_{om}^\prime F_- + \gamma_{om}^\prime F_R)} {(\gamma^2+\nu^2) (\gamma_{om}^{\prime 2}+\omega^2)}\right\} \frac{4C}{\gamma n_{x}} ~.
\end{aligned}
\end{equation}
Note the similarity between the first term and the expression for $\mathcal{S}_{\delta\hat{x}_1^\phi}^{SS}[\omega]$. The third term is a negative term that gives rise to the destructive interference. Fig. \ref{fig:phasenoise_shot} (b) plots $\mathcal{S}_{\delta\hat{n}_1^\phi}^{SS}[\omega]$ at different $\omega/\gamma$ and $C$. In the calculation $\gamma/\Omega=10^{-4}$ is assumed. Dips can be clearly observed at the corner of the ``L-shape'' spectrum. This destructive interference is more prominent in the USR due to the larger $\nu_{om}^\prime$. Another feature of the phase noise spectrum is that for the RSR the noise background decays when $\omega/\kappa>1$ representing the cavity filtering effect.

\begin{figure}
\centering
\includegraphics[width=8.6cm]{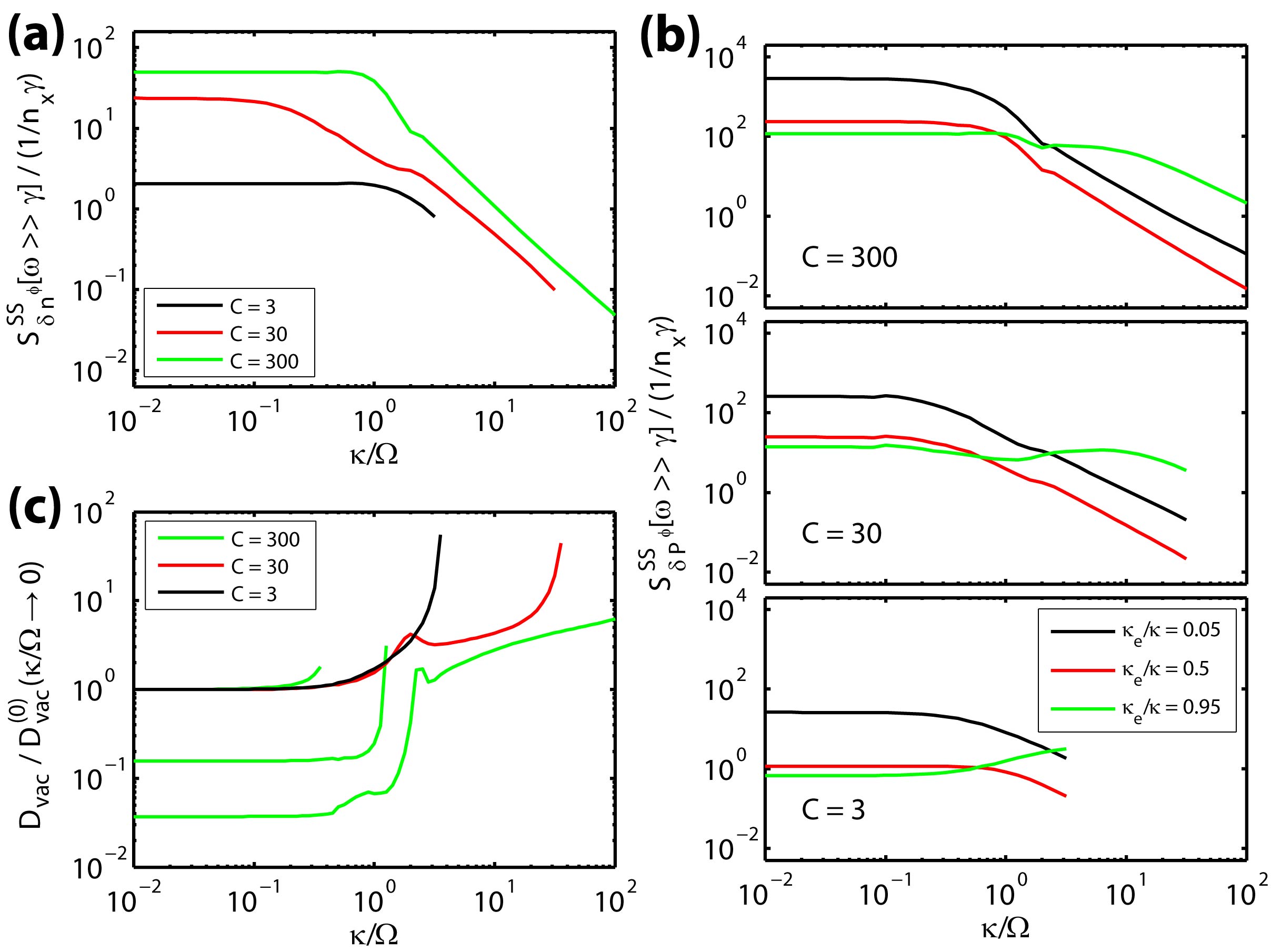}
\caption{ Noise floor of (a) $\mathcal{S}_{\delta\hat{n}_1^\phi}^{SS}$ and (b) $\mathcal{S}_{\delta\hat{P}_{out,1}^\phi}^{SS}$ plotted against $\kappa/\Omega$. (c) Normalized phase diffusion constant $\mathcal{D}_{vac} / \mathcal{D}_{vac}^{(0)} (\kappa/\Omega\rightarrow 0)$ plotted against $\kappa/\Omega$. $\mathcal{D}_{vac}^{(0)}$ denotes the diffusion constant for the limit-cycle with the smallest displacement when multiple solutions exist. }
\label{fig:Svc_floor}
\end{figure}

For oscillator applications, the noise background is another important figure-of-merit. This noise background term can be obtained from Eq. \eqref{eq:Snn_sh} by taking $\omega\gg\gamma$. It can be shown that
\begin{equation}\label{eq:Snn_sh_bg}
\begin{aligned}
\mathcal{S}_{\delta\hat{n}_1^\phi}^{SS} [\omega\gg\gamma] = \frac{\gamma^4 F_+ + \gamma^2\nu^2F_- + 2\gamma^3\nu F_R}{(\gamma^2+\nu^2)^2} \frac{4C}{\gamma n_x} ~.
\end{aligned}
\end{equation}
Fig. \ref{fig:Svc_floor} (a) plots the noise floor normalized with $1/n_x\gamma$ for the adiabatic region $\omega<\kappa$. It shows that the noise floor is generally lower in the USR.

For the phase noise of the output power $\delta\hat{P}_{out,1}$, the analytic expression is too lengthy to be included in the text here. We only show the numerical calculation results in Fig. \ref{fig:phasenoise_shot} (c), where the critical coupling condition $\kappa_e=\kappa/2$ is assumed. Compared to $\mathcal{S}_{\delta\hat{n}_1^\phi}^{SS}[\omega]$, $\mathcal{S}_{\delta\hat{P}_{out,1}^\phi}^{SS}[\omega]$ has a higher noise floor and the features of the destructive interference are not visible due to the extra vacuum noise contribution (see Eq. \eqref{eq:adiab_dPout_A_phi}). Fig. \ref{fig:Svc_floor} (b) plots the noise floor of $\mathcal{S}_{\delta\hat{P}_{out,1}^\phi}^{SS}[\omega]$ for different coupling condition. Similar to Fig. \ref{fig:Svc_floor} (a), the noise floor is higher in the RSR than in the USR. Also, for the RSR under-coupling ($\kappa_e/\kappa<0.5$) is preferred while for the USR over-coupling ($\kappa_e/\kappa>0.5$) is preferred.

Finally, we write down the expression for the phase diffusion constant due to photon shot noise
\begin{equation}\label{eq:Dsh}
\mathcal{D}_{vac} = \gamma\frac{2C}{n_x} \left[F_+ +\frac{\nu_{om}^{\prime2}}{\gamma_{om}^{\prime2}} F_- +\frac{\nu_{om}^{\prime}}{\gamma_{om}^{\prime}} 2F_R\right] ~.
\end{equation}
If the cross-correlation term $F_R$ is ignored, it reduces to the expression presented in Ref. \cite{Armour_PRL_2010}. Fig. \ref{fig:Svc_floor} (c) shows $\mathcal{D}_{vac}$ for different situations. Similar to Fig. \ref{fig:phasenoise_thermal} (d), for each value of $C$, the plotted value is normalized with the diffusion constant calculated at the limit of $\kappa/\Omega\rightarrow0$.

\subsection{Phase noise contribution from low-frequency technical laser noise}\label{subsec:phasenoise_tech}

The noise contribution from thermomechanical fluctuation and the photon shot noise discussed in the previous two sections set the fundamental limit on the lowest phase noise one can achieve. In practice, there are always additional noise sources that influence the system. For example, lasers are never quantum noise limited in reality. A common type of technical laser noise is a low-frequency noise with a characteristic $1/f^\alpha$ spectrum. Here we consider the simplest case where the technical laser noise is significant only at low-frequency within the range $|\omega|<\Omega/2$. In this case, we keep only the zeroth order component of the laser noise $\delta\hat{s}_{in,0}$ and denote this technical laser noise as $\delta\hat{s}_{in,0}=\delta\hat{s}_{tec}$.

\begin{figure}
\centering
\includegraphics[width=8.6cm]{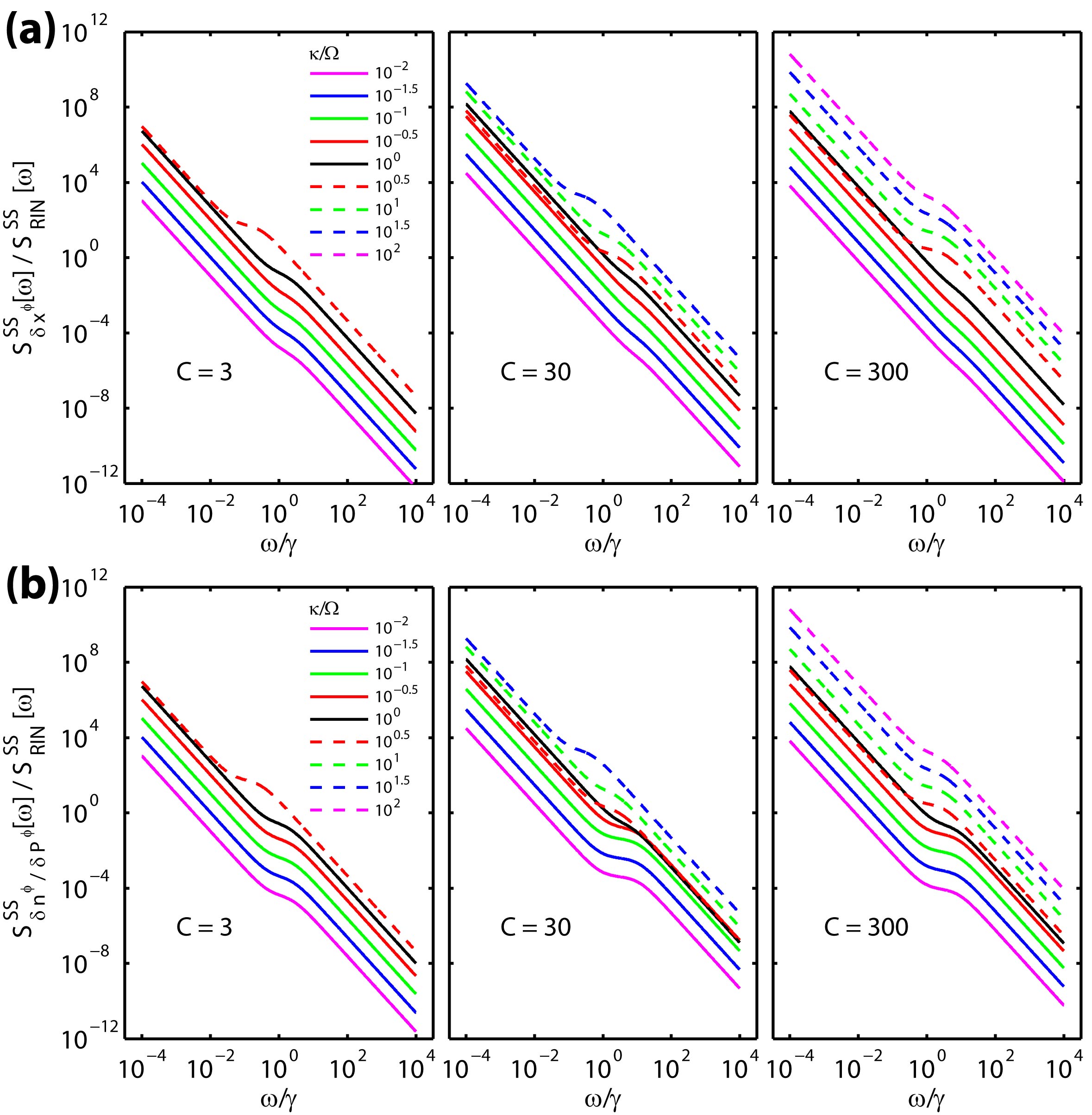}
\caption{ Phase noise spectral densities of (a) $\delta\hat{x}_1$ and (b) $\delta\hat{n}_1$ and $\delta\hat{P}_{out,1}$ due to low-frequency RIN noise. Solid lines are used for the RSR and dashed lines are used for the USR. }
\label{fig:phasenoise_tech}
\end{figure}

In the adiabatic limit when $L_{K_m}[\omega\rightarrow0]\rightarrow1$, the technical optical force noise has the following simple form,
\begin{equation}\label{eq:dfop_dstc}
\begin{aligned}
\delta\hat{f}_{op}^A &= \frac{2x_1}{\sqrt{2\gamma}} \gamma \delta\hat{s}_{tec}^A \\
\delta\hat{f}_{op}^\phi &= \frac{-2x_1}{\sqrt{2\gamma}} \nu \delta\hat{s}_{tec}^A ~.
\end{aligned}
\end{equation}
Interestingly, the force noise only depends on laser amplitude noise but not the laser phase noise nor the amplitude-phase cross-correlations. It is actually expected since the optical force is given by $g\hat{a}^\dagger\hat{a}$ and so the phase does not come into play, unless at time scales shorter than the cavity response ($\omega>\kappa$) will the transient response of the cavity turn the laser phase noise into optical force noise. From Eq. \eqref{eq:dfop_dstc} we can see that for device with large $\nu/\gamma$ ratio the force noise is mostly acting along the phase direction. It can be shown that the force noise correlation matrix is given by
\begin{equation}\label{eq:corr_matrix_ftc}
\mathcal{C}_{\delta\hat{f}_{tec}}[\omega]= \frac{2n_x}{\gamma} \mathcal{S}_{\delta\hat{s}_{tec}^A \delta\hat{s}_{tec}^A}[\omega]
\left[\begin{array}{cc}
\gamma^2 & -\gamma\nu \\
-\gamma\nu & \nu^2
\end{array}\right] ~.
\end{equation}
For phase noise spectral densities, it is useful to express it in terms of the laser relative intensity noise (RIN), which is a commonly used quantity specifying laser noise. The relative intensity noise is given by $\delta\hat{P}/P=2\delta\hat{s}^A$ and therefore $\mathcal{S}_{\mathrm{RIN}}^{SS}[\omega] =4\mathcal{S}_{\delta\hat{s}^A}^{SS}[\omega]$. Using this notation, the phase noise spectral density for $\delta\hat{x}_1$ is
\begin{equation}\label{eq:Sxx_tc}
\mathcal{S}_{\delta\hat{x}_1^\phi}^{SS} [\omega] = \frac{\eta_I^2(\gamma^2+\nu^2)^2 +\nu^2\omega^2} {\omega^2(\gamma_{om}^{\prime2}+\omega^2)} \mathcal{S}_{\mathrm{RIN}}^{SS} [\omega] ~.
\end{equation}
For $\delta\hat{n}_1$ and $\delta\hat{P}_{out,1}$, it can be shown that they have the same expression and are given by
\begin{equation}\label{eq:Snn_tc}
\begin{aligned}
&\mathcal{S}_{\delta\hat{n}_1^\phi}^{SS} [\omega] =\mathcal{S}_{\delta\hat{P}_{out,1}^\phi}^{SS} [\omega] \\
&=\frac{\eta_I^2(\gamma^2+\nu^2)^2 +(\eta_I\gamma-\nu)^2\omega^2} {\omega^2(\gamma_{om}^{\prime2}+\omega^2)} \mathcal{S}_{\mathrm{RIN}}^{SS}[\omega] ~.
\end{aligned}
\end{equation}
Note that this time the phase noise spectrum is a decreasing function with no flat background, since the background term (second term in Eq. \eqref{eq:adiab_dn_A_phi}) is acting all in the amplitude direction but not the phase direction. Fig. \ref{fig:phasenoise_tech} (a) and (b) plot the transfer function for different $\kappa\Omega$ and $C$. In all cases, a general trend is that the transfer function is higher in the USR due to the large mechanical detuning $\nu$, which leads to larger force noise along the phase direction.

\subsection{Sample calculations}\label{subsec:sample_calc}

\begin{figure}
\centering
\includegraphics[width=8.6cm]{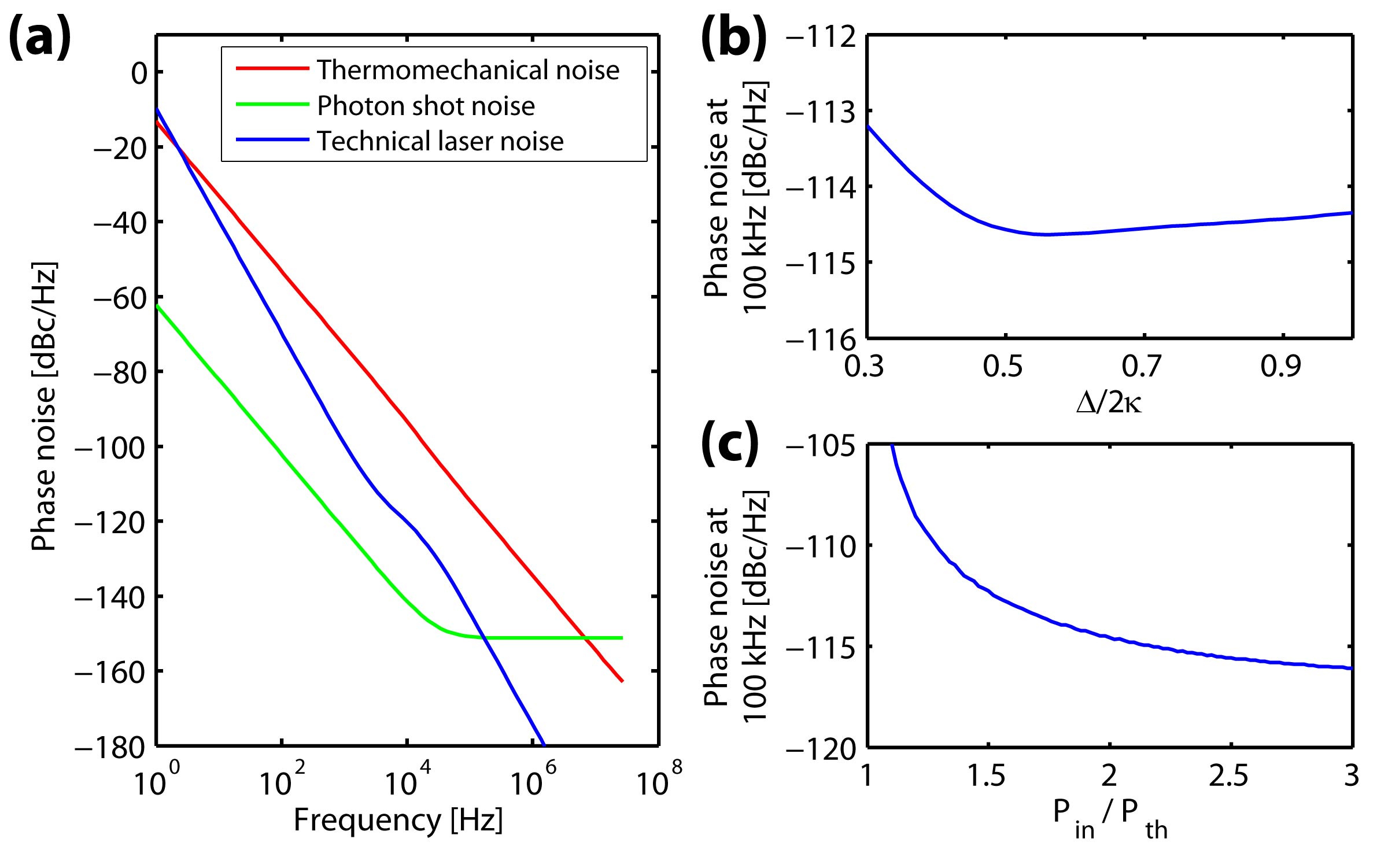}
\caption{ (a) Phase noise spectrum calculated for the optomechanical oscillator reported in Ref. \cite{Vahala_PRA_2006}. (b) Phase noise at 100 kHz plotted against $\Delta/2\kappa$. (c) Phase noise at 100 kHz plotted against $P_{in}/P_{th}$. (a), (b) and (c) are for comparison with Figs. 7, 9 (a) and 9 (b) of Ref. \cite{Vahala_PRA_2006}. }
\label{fig:sample_calc}
\end{figure}

To apply the phase noise theory presented here, we calculate the phase noise spectrum for the first demonstrated optomechanical oscillator reported in Refs. \cite{Vahala_OE_2005, Vahala_PRA_2006, Vahala_APL_2006}. The demonstrated device is a high Q silica micro-toroid optical resonator. A detailed characterization of this device can be found in Ref. \cite{Vahala_PRA_2006}. Here we list the experimentally measured parameters of the device in table \ref{tab:table1}. In the calculation, a $1/f$ laser RIN with a value of -120 dBc/Hz at 10 kHz offset frequency is assumed. Fig. \ref{fig:sample_calc} (a) plots the calculated phase noise (defined as $\mathcal{L}[\omega] = \mathcal{S}_{\delta\hat{P}^\phi}^{SS}[\omega]/2$). We can see that in most of the frequency range the device phase noise is dominated by the thermomechanical noise. The contribution from the thermomechanical noise is almost 50 dB higher than that from the photon shot noise in the $1/f^2$ regime. The noise floor imposed by the photon shot noise is at around -150 dBc/Hz. It was not observed in the original experiment probably due to the higher instrument noise background. For the contribution from technical laser noise, its effect is negligible compared with that from thermomechanical noise unless the offset frequency is below 10 Hz. Therefore, the $1/f^3$ noise observed in the original experiment is likely to be due to other noise source.

Besides the full phase noise spectrum, we also calculate the phase noise at 100 kHz offset frequency at different detunings and input powers, as shown in Fig. \ref{fig:sample_calc} (b) and (c). These two figures are to be compared with the Fig. 9 (a) and (b) in Ref. \cite{Vahala_PRA_2006}. The calculated values and the experimental results show qualitative agreement.

\begin{table}[b]
\caption{\label{tab:table1}
List of parameters for device reported in Ref. \cite{Vahala_PRA_2006}}
\begin{ruledtabular}
\begin{tabular}{lll}
Parameters & Symbols & Values \\
\colrule
Mechanical frequency         & $\Omega_m/2\pi$ & 54.2 MHz \\
Mechanical $Q$               & $Q_m$           & 2100 \\
Effective mass               & $m_e$           & 23 ng \\
Laser wavelength             & $\lambda$       & 1550 nm \\
Intrinsic optical $Q$        & $Q_i$           & $5.5\times10^6$ \\
Loaded optical $Q$           & $Q_o$           & $1.5\times10^6$ \\
Cavity dissipation rate      & $\kappa/2\pi$   & 64.5 MHz \\
Input coupling rate          & $\kappa_e/2\pi$ & 46.9 MHz \\
Optical detuning             & $\Delta/2\kappa$& 0.5 \\
Optomechanical coupling rate & $g/2\pi$        & 245 Hz\footnote{Estimated from the reported oscillation threshold power}
\end{tabular}
\end{ruledtabular}
\end{table}

\subsection{Comparison with the Leeson's model}\label{subsec:leeson_model}

\begin{figure}
\centering
\includegraphics[width=8.6cm]{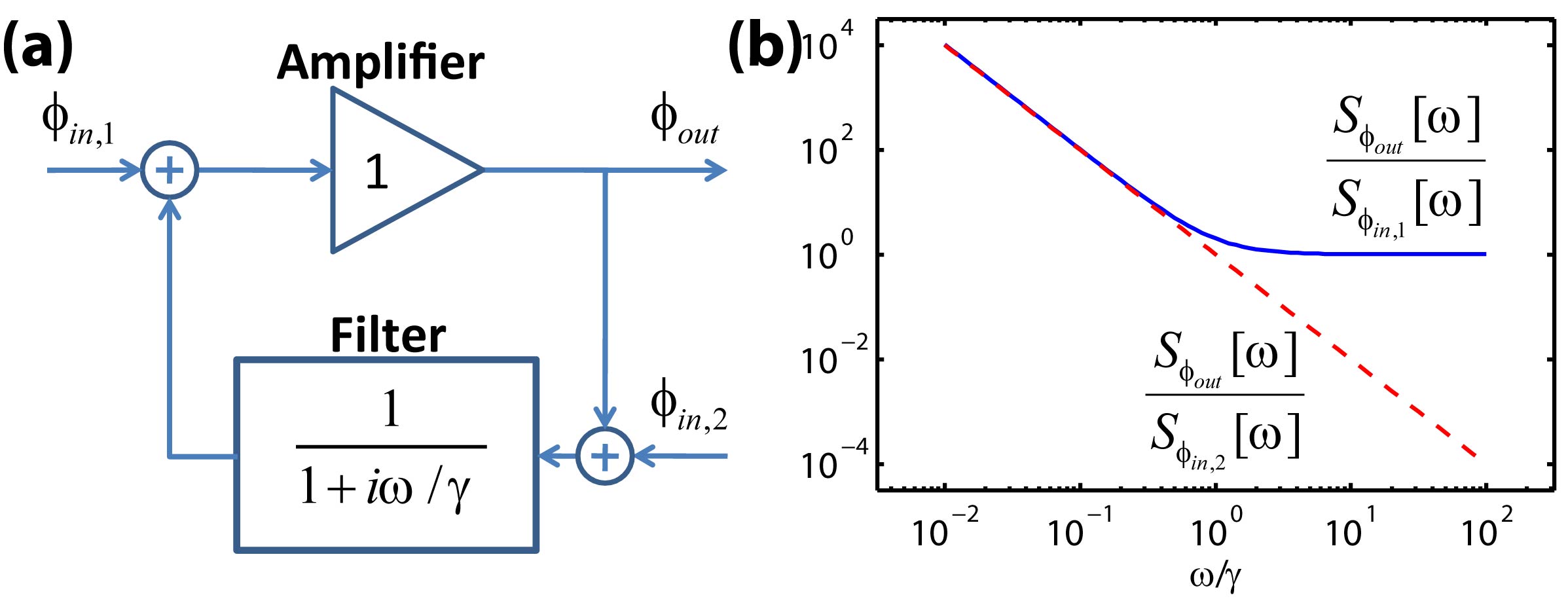}
\caption{ (a) Simplified oscillator feedback loop in the phase noise space. (b) Phase noise transfer function with respect to input 1 and input 2. }
\label{fig:leeson_model}
\end{figure}

A final remark about the phase noise analysis presented here is that it is fully compatible with the well-known Leeson's model \cite{Leeson_IEEE_1966}, which is a widely used model in the study of the phase noise of electronic oscillators. 
For comparison, here we briefly describe the derivation of the Leeson's formula following the transfer function approach presented in Ref. \cite{Rubiola_Book_2010} and compare it with the analysis presented here. Fig. \ref{fig:leeson_model} (a) shows the simplified oscillator feedback loop in the phase noise space. The amplifier has a gain of 1 and the resonator has a transfer function of $1/(1-i\omega/\gamma)$ (the resonator is assumed to be driven at resonance). It can be shown that the phase noise transfer function with respect to the input at the amplifier (input 1) is given by $\mathcal{S}_{\phi_{out}}[\omega] /\mathcal{S}_{\phi_{in,1}}[\omega] = 1 + \gamma^2/\omega^2$ and the phase noise transfer function with respect to the input at the resonator (input 2) is given by $\mathcal{S}_{\phi_{out}}[\omega] /\mathcal{S}_{\phi_{in,2}}[\omega] = \gamma^2/\omega^2$, as plotted in Fig. \ref{fig:leeson_model} (b). The expression for the phase noise contributed from the noise at the input 1, i.e., $1 + \gamma^2/\omega^2$, is the famous Leeson's formula.

It can be shown that the phase noise analysis for the optomechanical oscillators presented in this manuscript can be reduced to the same form if all the amplitude noise contributions are ignored and only the phase noise terms are kept. For example, for the phase noise of $\delta\hat{n}_1^\phi$ described in Eq. \eqref{eq:adiab_dn_A_phi}, if we keep only the phase noise components (keep only the lower-right matrix element), assume zero mechanical detuning $\nu=0$ and take the normalization factor of $\sqrt{2/x_1^2\gamma}$ to be 1, we can immediately see that the equation becomes $\delta\hat{n}_1^\phi=\frac{i\gamma}{\omega} \delta\hat{f}_{th}^\phi + (1+\frac{i\gamma}{\omega}) \delta\hat{f}_{op}^\phi$, which leads to the same expression for the noise spectral density as the Leeson's formula.

\section{Conclusion}\label{sec:conclusion}

In conclusion, we theoretically analyzed the phase noise of a self-oscillating cavity optomechanical system. We derived expressions for the transfer functions for the optical cavity and the mechanical resonator, which can be considered as the two components of the oscillator feedback loop. The transfer functions of each component were combined to solve for the noise response of the closed-loop system. Expressions for the phase noise spectral densities contributions from thermomechanical noise, photon shot noise, and low-frequency technical laser noise were derived. We numerically calculated the phase noise for an experimentally demonstrated system, which agrees qualitatively with the experimental results. We also showed that the presented model reduces to the form of the well-known Leeson's model of phase noise when amplitude noise is ignored.


\begin{acknowledgments}
M.P. thanks the Netherlands Organization for Scientific Research (NWO)/Marie Curie Cofund Action for support via a Rubicon fellowship. H.X.T. acknowledges support from a Packard Fellowship in Science and Engineering and a career award from National Science Foundation. This work was funded by the DARPA/MTO ORCHID program through a grant from the Air Force Office of Scientific Research (AFOSR).
\end{acknowledgments}

\appendix

\section{Spectral filter decomposition}\label{app:spectral_decomposition}

We define the rectangular window function as
\begin{equation}\label{eq:A1}
\mathrm{rect}(x)=\left\{
\begin{array}{ll}
0, & x\le-\frac{1}{2} \\
1, & -\frac{1}{2}<x\le\frac{1}{2} \\
0, & x>\frac{1}{2}
\end{array}
\right. ~.
\end{equation}
This definition ensures that there is no overlap between $\mathrm{rect}(x)$ and $\mathrm{rect}(x\pm 1)$. Then we have the following identity holds for all real number $\omega$,
\begin{equation}\label{eq:rect_identity}
\sum_n \mathrm{rect} \left( \frac{\omega-n \Omega}{\Omega} \right) = 1 ~.
\end{equation}
For any operator that is function of time $\hat{f}(t)$, its inverse Fourier transform is given by
\begin{equation}
\hat{f}(t)=\frac{1}{2\pi}\int_{-\infty}^{\infty}{d\omega
\hat{f}[\omega]e^{-i\omega t}} ~.
\end{equation}
By inserting the identity of Eq. \eqref{eq:rect_identity} into the integral and using a change of variable $\omega\rightarrow\omega-n\Omega$, the Fourier transform can be written in the form of
\begin{equation}
\hat{f}(t)=\sum_n{\hat{f}_n(t)e^{-in\omega t}} ~,
\end{equation}
where the harmonic components $\hat{f}_n$ in frequency domain are given by
\begin{equation}
\hat{f}_n[\omega]= \hat{f}[\omega+n\Omega]\mathrm{rect}(\omega/\Omega) ~.
\end{equation}
Therefore, the frequency spectrum of $\hat{f}_n(t)$ is nonzero only for $-\Omega/2<\omega\le\Omega/2$. It can be further shown that the cross power spectral densities of $\hat{f}_n(t)$ are given by
\begin{equation}\label{eq:Math:WSD_PSD}
\begin{aligned}
\mathcal{S}_{\hat{f}_n \hat{f}_m}[\omega] &= \mathcal{S}_{\hat{f}\hat{f}}[\omega+n\Omega] \mathrm{rect}(\omega/\Omega) \delta_{n,-m} \\
\mathcal{S}_{\hat{f}_n \hat{f}_m^\dagger}[\omega] &= \mathcal{S}_{\hat{f}\hat{f}^\dagger}[\omega+n\Omega] \mathrm{rect}(\omega/\Omega) \delta_{n,m} \\
\mathcal{S}_{\hat{f}_n^\dagger \hat{f}_m}[\omega] &= \mathcal{S}_{\hat{f}^\dagger\hat{f}}[\omega-n\Omega] \mathrm{rect}(\omega/\Omega) \delta_{n,m} \\
\mathcal{S}_{\hat{f}_n^\dagger \hat{f}_m^\dagger}[\omega] &= \mathcal{S}_{\hat{f}^\dagger\hat{f}^\dagger}[\omega-n\Omega] \mathrm{rect}(\omega/\Omega) \delta_{n,-m} ~.
\end{aligned}
\end{equation}

\bibliography{preprint_bib}

\end{document}